\documentclass[12pt,onecolumn,manuscript,tighten,trackchanges]{aastex62}
\usepackage{natbib}
\usepackage{multirow,color}
\usepackage{bbding}
\usepackage{amssymb}
\usepackage{ulem}
\usepackage{color}
\usepackage{longtable}
\usepackage{array}
\usepackage{bm}
\usepackage{url}
\usepackage{CJKutf8}
\usepackage{graphicx}
\usepackage{enumerate}

\def\gsim{\;\lower4pt\hbox{${\buildrel\displaystyle >\over\sim}$}\;}
\def\lsim{\;\lower4pt\hbox{${\buildrel\displaystyle <\over\sim}$}\;}
\def\grls{\;\lower4pt\hbox{${\buildrel\displaystyle >\over <}$}\;}

\bibpunct{(}{)}{,}{o}{}{,} 

\begin{document}

\title{Formation of a Magnetic Flux Rope in the Early Emergence Phase of NOAA Active Region 12673} 

\author{Lijuan Liu
\begin{CJK}{UTF8}{gbsn}
(刘丽娟)
\end{CJK}}
\affiliation{School of Atmospheric Sciences, Sun Yat-sen University, Zhuhai, Guangdong, 519082, China}
\affiliation{CAS center for Excellence in Comparative Planetology, China}
\affiliation{School of Astronomy and Space Science, Nanjing University, Nanjing, 210023, China}
\email{liulj8@mail.sysu.edu.cn}

\author{Xin Cheng
\begin{CJK}{UTF8}{gbsn}
(程鑫)
\end{CJK}}
\affiliation{School of Astronomy and Space Science, Nanjing University, Nanjing, 210023, China}
\affiliation{Max Planck Institute for Solar System Research, Gottingen, 37077, Germany}
\email{xincheng@nju.edu.cn}

\author{Yuming Wang
\begin{CJK}{UTF8}{gbsn}
(汪毓明)
\end{CJK}}
\affiliation{CAS Key Laboratory of Geospace Environment, Department of Geophysics and Planetary Sciences, University of Science and Technology of China, Hefei, Anhui, 230026, China}
\affiliation{CAS center for Excellence in Comparative Planetology, China}

\author{Zhenjun Zhou
\begin{CJK}{UTF8}{gbsn}
(周振军)
\end{CJK}}
\affiliation{School of Atmospheric Sciences, Sun Yat-sen University, Zhuhai, Guangdong, 519082, China}
\affiliation{CAS center for Excellence in Comparative Planetology, China}

\begin{abstract}

In this work, we investigate the formation of a magnetic flux rope (MFR) above the central polarity inversion line (PIL) of NOAA Active Region 12673 during its early emergence phase. Through analyzing the photospheric vector magnetic field, extreme ultraviolet (EUV) and ultraviolet (UV) images, 
extrapolated three-dimensional (3D) non-linear force-free fields (NLFFFs), as well as the photospheric motions, 
we find that with the successive emergence of different bipoles in the central region, 
the conjugate polarities separate, 
resulting in collision between the non-conjugated opposite polarities. 
Nearly-potential loops appear above the PIL at first, then get sheared and merge at the collision locations 
as evidenced by the appearance of a continuous EUV sigmoid 
on 2017 September 4, 
which also indicates the formation of an MFR. 
The 3D NLFFFs further reveal the gradual buildup of the MFR, 
accompanied by the appearance of two elongated bald patches (BPs) at the collision locations and a very low-lying hyperbolic flux tube configuration between the BPs. 
The final MFR has relatively steady axial flux and average twist number of around $2.1\times 10^{20}$~Mx and -1.5, respective. 
Shearing motions are found developing near the BPs when the collision occurs, 
with flux cancellation and UV brightenings being observed simultaneously, 
indicating the development of a process named as ``collisional shearing"~\citep[firstly identified by][]{Chintzoglou_2018}.  
The results clearly show that the MFR is formed by ``collisional shearing", i.e., through shearing and flux cancellation driven by the collision between non-conjugated opposite polarities during their emergence. 

\end{abstract}

\section{INTRODUCTION}\label{sec:intro}

A magnetic flux rope (MFR) is a collection of magnetic field lines wrapping around a common axis with at least one full turn. 
It is a fundamental structure in the magnetized solar atmosphere~\citep[e.g.,][and references therein]{Chenj_2017, Chengx_2017}. 
Solar flare and Coronal mass ejection (CME), the most spectacular eruptions happening in the solar atmosphere, are believed to be two 
manifestations of the eruption of an MFR as interpreted by the standard flare model~\citep[e.g.][]{Shibata_1995}. 
After the MFR erupts, it quickly forms a CME, 
accompanied with a flare generated by the magnetic reconnection at the current sheet in the wake of the erupting CME. 
The former releases bulk of plasma and magnetic flux into the interplanetary space, 
and the latter causes strong emissions in various wavebands, 
both having potential to induce hazardous space weather in the geospace.

Although the MFR is believed to be the fundamental structure of the solar eruption, 
it is still under debate that whether the MFR exists prior to the eruption. 
In breakout and tether-cutting models, 
the pre-eruption structure is considered to be sheared arcades which are transformed into the MFR through reconnection during the eruption~\citep{Antiochos_1999, Moore_2001}. 
In contrast, 
pre-eruption MFR is required by the ideal MHD instabilities, e.g., kink instability~\citep{Torok_2004} and torus instability~\citep{Kliem_2006}.

To distinguish the two kinds of the models, many studies paid attention to the origin of the pre-eruption configuration, 
in particular the MFR.  
At present, two general types of formation mechanisms for pre-eruption MFR are proposed: 
bodily emergence from below the photosphere or direct formation in the solar atmosphere. 
In the first scenario, the MFR is believed to be formed in the convection zone and emerges into the solar atmosphere through magnetic buoyancy \citep{Zwaan_1985, Blow_2001}. 
The process is investigated by many numerical simulations. 
It is found that the bodily emergence of the MFR, 
i.e., the emergence of its axis and the helical field lines around it,  
can only be achieved by imposing strong field strength or high degree of twist to the subsurface MFR, or vertical velocity field in the convection zone~\citep[e.g.,][]{Emonet_1998, Fan_2004, Amari_2004, Amari_2005, Murray_2006, Jouve_2009}. 
In most cases, 
the MFR axis is not able to cross through the photosphere by the means of magnetic buoyancy and magnetic buoyancy instability,  
since the heavy plasma collected at the magnetic dips inhibits its emergence~\citep[][and reference therein]{Cheung_2014}. 
The emerged part appears as sheared arcades in the solar atmosphere. 
It is the subsequent photospheric flows and flows-driven reconnection that play a crucial role in creating a new MFR in the solar atmosphere~\citep[e.g.,][]{Fanyh_2001, Archontis_2004, ManchesterIV_2004, Archontis_2008a, Fanyh_2009, Archontis_2010, Archontis_2012}. 
The process is similar to a ``serpentine flux tube" emergence scenario that interprets the origin of the fine-scale coronal magnetic field~\citep[e.g.,][]{Bernasconi_2002, Pariat_2004, Cheung_2008, Pariat_2009a,  Valori_2012, Cheung_2014}.

The bodily emergence scenario is also supported by a few observations. 
For example,~\cite{Okamoto_2008, Okamoto_2009} examined the vector magnetograms of NOAA active region (AR) 10593 and found two primary phenomena based on which they concluded 
the bodily emergence of a horizontal MFR: 
(1) the abut opposite-polarity regions with weak vertical magnetic field and strong horizontal magnetic field first grow laterally then narrow down  (named as ``sliding-door" effect), 
(2) the horizontal field along the polarity inversion line (PIL) reverse from normal-polarity (pointing from positive to negative polarity) to inverse-polarity configuration. 
Moreover, blue-shift, ``sliding-door" effect, etc., in filament regions are reported to suggest possible bodily emergence of the MFR~\citep[e.g.,][]{Lites_1995, Lites_2010, Kuckein_2012a, Kuckein_2012b}. 
However, the results in~\cite{Okamoto_2008} are also questioned by some other researchers, e.g.,~\citet{Vargas_2012}, 
due to the lack of the prominent flux emergence and characteristic flows. 
Furthermore, statistical works suggested that more than $90\%$ of intermediate and quiescent filaments, which are seen as the MFR proxy, 
appear in the region involving interaction of multiple bipoles, 
rather than in the single bipolar region, suggesting that bodily emergence of the MFR may be rare~\citep[][and reference therein]{Mackay_2010}.

In the second scenario, the MFR is thought to be directly formed in the solar atmosphere via magnetic reconnection and/or photospheric motions. 
For instance, in the flux cancellation model proposed by~\citet{VanBallegooijen_1989}, 
through photospheric shearing motions and converging motions in a bipolar region, 
the coronal arcades are sheared and brought together to reconnect at, or slightly above, the photosphere, 
forming the helical flux of the MFR as well as the small magnetic loops underneath. 
If the loops are short and low enough~\citep[lower than a few times of the photosphere scale height, 
i.e., around several tenths of a megameter from the base of the photosphere as indicated in][]{VanBallegooijen_1989},  
they may submerge due to strong magnetic tension, manifesting as flux cancellation at the photosphere. 
The process has been successfully reproduced 
by numerical simulations~\citep[e.g.,][]{Amari_2003a, Amari_2003b, Aulanier_torok_2010, Amari_2010a, Jiang_2014a} 
in which various surface effects 
such as shearing motions, converging motions, turbulent diffusion, etc., are employed 
to drive the flux cancellation. 
As an extension to this model, 
several works suggested that the flux cancellation may occur at the interface of different bipoles within a multipolar region~\citep{Martens_2002, Welsch_2005, Mackay_2006}.

Observationally, the MFR formation through flux cancellation has also been widely investigated~\citep[e.g.,][]{Schmieder_2004b, Green_2009, Green_2011, Savcheva_2012, Cheng_2014a, Joshi_2014a, Yanxl_2016}. 
As the direct measurements of the coronal magnetic field are quiet rare, 
various proxies of the MFR, e.g., filaments, sigmoids, hot channels are investigated to reveal the MFR formation. 
For example,~\cite{Green_2009} and \cite{Green_2011} reported the transition from 
two groups of J-shaped sheared arcades into a continuous sigmoid in the soft X-ray emission, 
similar to the formation of a sigmoidal hot channel in the 94~\AA~and 131~\AA~passbands as reported by~\cite{Cheng_2014a}. 
In these cases, 
the flux cancellation is observed, 
with a unique configuration called Bald Patch~\citep[BP,][]{Titov_1993} 
identified. 
Magnetic field lines at BP graze the photosphere and point from the negative to the positive polarity, that the separatrix surface favoring the reconnection may be formed here. 
Considering the presence of prominent shearing flows along, and converging flows toward the PIL, 
the authors suggested that the flux cancellation driven by the photospheric flows near the BP leads to the formation of the MFRs.

The flux cancellation is indeed a reconnection happening at (or very near to) the photosphere and usually proceeds gradually in a quasi-equilibrium manner.  
It is possible for the pre-eruption MFR to be formed by reconnection higher (in the corona), or more drastically (in confined flares). 
With the slow rise of the forming MFR, 
the photospheric BP configuration may change to a coronal Hyperbolic Flux Tube~\citep[HFT,][]{Titov_2002} topology, at which two quasi-separatrix layers~\citep[QSL,][]{Titov_2002} intersect. 
The reconnection may occur here to from the MFR~\citep[e.g.,][]{Aulanier_torok_2010, Fan_2012, Cheng_2014a}. 
The process may even occur rapidly during confined flares as 
suggested by a few observations~\citep[e.g.,][]{Guoy_2013, Patsourakos_2013, Chintzoglou_2015, James_2017, Liu_2018d}. 
Moreover, the MFR may also be formed by photospheric motions without reconnection. 
For example, 
sunspot rotation is suggested to be able to twist the magnetic field lines to form the MFR~\citep[e.g.,][]{Longcope_2000, Fanyh_2009, Leakej_2013, Yanxl_2015, Yanxl_2018a}. 

As described above, 
the photospheric flows and flows-driven magnetic reconnection 
play an important role in forming the pre-eruption MFR in the solar atmosphere, 
either during the partial emergence of a sub-surface MFR or in the process not considering the sub-surface magnetic configuration. 
Various flows may be caused by various mechanisms. 
For instance, the shearing flows may result from differential rotation~\citep{Devore_2000, Welsch_2005} or driven by the magnetic tension~\citep{ManchesterIV_2004}; the converging flows can be caused by the flux diffusion~\citep{Aulanier_torok_2010}; the rotation motions may result from the propagation of the nonlinear torsional Alv\'en wave along  the flux tube~\citep{Longcope_2000, Fanyh_2009}. 
Recently,~\cite{Chintzoglou_2018} proposed a new interpretation to the causes of the photospheric shearing flows, 
as well as the origin of major solar activities, through analyzing two flare/CME productive Active Regions (ARs). 
The ARs have multipolar configuration formed by the multiple flux tubes emerging simultaneously or sequentially. 
The two legs of each flux tube, which are manifested as the conjugated polarities on the photosphere, naturally separate during their emergence, resulting in collision between non-conjugate opposite polarities~\citep[see Figure 1 in][]{Chintzoglou_2018}. 
The collision drives subsequent shearing and cancellation that produce the activities. 
The process is named as ``collisional shearing".   

In short, it seems that for the pre-eruption MFR, being formed in the solar atmosphere is more likely than its bodily emergence.  
If the latter does occur, flux emergence would be a natural result. 
Meanwhile, observationally, flux emergence is usually accompanied with various surface effects such as photospheric motions and flux cancellation.   
Due to their mixture, 
disclosing the exact origin of the MFR during the flux emergence becomes difficult.  
In this paper, we investigate the formation of an MFR in the early emergence phase of NOAA AR 12673 to explore the above problem. 
The AR is the most productive AR in the minimum of Solar Cycle (SC) 24 and presents the fastest flux emergence ever observed~\citep{Sunxd_rna_2017}, 
hosting so far the most energetic flare (an X9.3 class flare on 2017 September 6) in SC 24. 
Most research focused on the properties of the two X-class flares on 2017 September 6~\citep[e.g.,][]{Yangsh_2017, Romano_2018, Mitra_2018a, Inoue_2018a, Verma_2018, Yanxl_2018a, Veronig_2018, Shen_2018, Hou_2018, Zou_2019}.  
A few work also paid attention to the long-term evolution of the AR. 
For instance,~\cite{Wang_2018e} analyzed the photospheric flows and concluded that the horizontal flows on the photosphere contributed the majority of the magnetic helicity of the AR. 
\cite{Vemareddy_2019} analyzed the extrapolated coronal magnetic field in addition to the photospheric flows and drew a similar conclusion. 
Although both works hint that the MFR may be formed in the solar atmosphere rather than from a bodily emergence, 
a detailed study on the formation process at the central PIL (the source PIL of the two largest X-class flares), especially during its early emergence phase, is still needed. 
The remainder of the paper is organized as follows: 
the data and methods are introduced in Section~\ref{sec:data}, 
the results are presented in Section~\ref{sec:res}, 
followed by the summary and discussions in Section~\ref{sec:conc}.

\section{DATA AND METHODS}\label{sec:data}

To analyze the origin of the MFR above the central PIL of NOAA AR 12673, 
the evolution of the photospheric magnetic field and coronal appearance in ultraviolet (UV) and extreme ultraviolet (EUV) passbands are inspected at first. 
Data from the HMI~\citep{Hoeksema_etal_2014} and the AIA~\citep{Lemen_etal_2012} onboard {\it SDO}~\citep{Pesnell_2012} is mainly used. 
The HMI provides a data product called {\it Space-weather HMI Active Region Patches} (SHARPs)~\citep{Bobra_2014}, 
which tracks the photospheric vector field 
of the AR with a spatial resolution of 0.5$''$ and a temporal cadence of 12 minutes. 
A specific version of the SHARP data, which is remapped into the cylindrical equal area (CEA) coordinate to correct the projection effect, 
is used. The AIA provides the UV and EUV images with a temporal cadence up to 12 seconds and a spatial resolution of 0.6$''$.

Due to the lack of direct measurement, the three-dimensional (3D) coronal magnetic field is reconstructed by a Non-linear force-free fields (NLFFFs) extrapolation model~\citep{Wiegelmann_2004, Wiegelmann_2006, Wiegelmann_2012}, 
using the CEA version of SHARP magnetograms as the bottom boundaries. 
The magnetograms are preprocessed to reconcile the possible non-force-freeness of the photosphere and the force-free assumption of the model.  

Based on the extrapolated coronal magnetic field, we quantitatively analyze the property of the MFR. 
We identify the MFR strucutre using a method developed by~\cite{rliu_2016} through calculating the twist number $T_w$ and squashing factor $Q$ for the magnetic field lines. 
The former measures the number of turns a field line winding and is computed by $T_w = \frac{1}{4\pi}\int_l\alpha dl$, 
where $\alpha$ denotes the force free parameter and $dl$ denotes the elementary length. 
The latter quantifies the change of the connectivity of the magnetic field lines. High $Q$ values mark the QSL 
region, where the connectivity of the field lines changes drastically~\citep{Demoulin_1996, Demoulin_1997, Demoulin_2006}. 
An MFR has high degree of twist than its surrounding, 
that its cross section generally displays a strong twisted region wrapped by a boundary of extremely high Q values. 
Once the MFR is identified in the NLFFFs using the method above,   
its axial flux $\Phi$ and average twist number $\overline{T}_w$ can be calculated. $\Phi$ is computed by $\Phi=\bf{B}\cdot\bf{S}$, 
where ${\bf S}=S\bf{n}$; $S$ is the area of the MFR cross section 
while $\bf{n}$ is its normal unit vector. 
The average twist number $\overline{T}_w$ can be estimated by $\displaystyle \overline{T}_w=\frac{\Sigma T_w(|B_n|dA)^2}{\Sigma (|B_n|dA)^2}$ based on the expression of the relative magnetic helicity of a flux tube, 
$H=\int T_wd\phi^2\approx\overline{T}_w\int d\phi^2$~\citep{Berger_Field_1984}. 
$B_n$ is the magnetic field component normal to the cross section, and $dA$ is the elementary area.  

We also calculate the photospheric velocities by a Differential Affine Velocity Estimator for Vector Magnetograms (DAVE4VM) method developed by~\cite{Schuck_2008}, 
using a time series of SHARP magnetograms as input.

\section{RESULTS}\label{sec:res}

\subsection{Formation of the MFR along with the flux emergence}\label{subsec:formation} 
\subsubsection{Flux emergence in the central region}\label{subsubsec:emergence} 

\begin{figure*}
\begin{center}
\epsscale{1.1}
\plotone{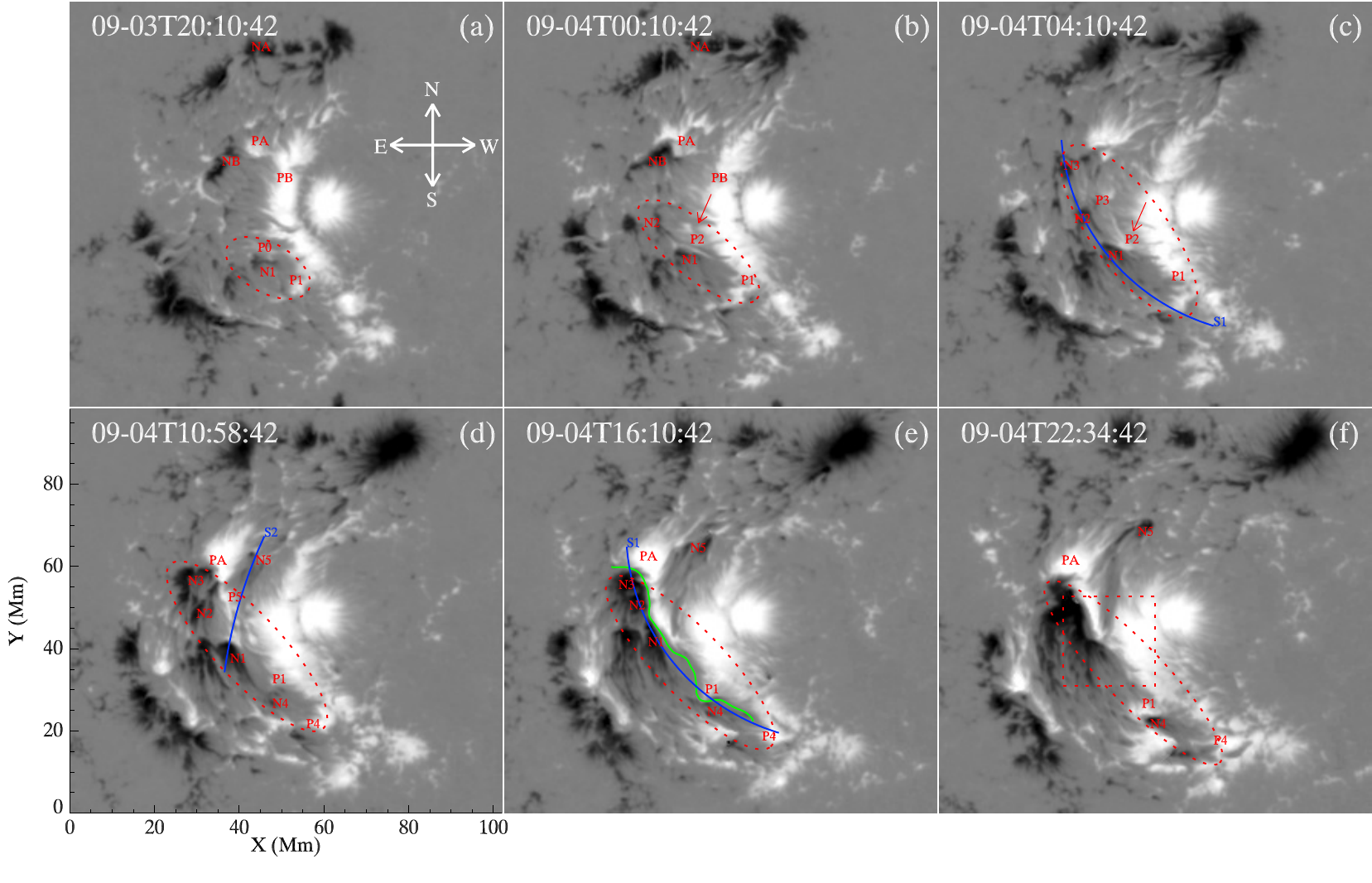}
\caption{Maps of vertical component ($B_z$) of the photospheric vector magnetic field at six selected moments, 
showing the emergence of the different flux concentrations forming the central PIL of the AR. The interested region 
is enclosed by the red ellipse.  
The field saturates at $\pm2000$ Gauss, with black (white) patches for negative (positive) polarities. 
Labels N1 P1, N2 P2, N3 P3, N4 P4 and N5 P5 denote the identified bipoles,  
N (P) for negative (positive) polarities. 
P0 denotes a pre-existing positive polarity next to N1. 
NA PA and NB PB denote two pre-existing bipoles.   
The red rectangle in panel (f) indicates 
the field of view (FOV) of Figure~\ref{fig:bz_cancel}. 
The red arrows in panel (b)-(c) point out the southward motion of P2. 
The blue arcs S1 and S2 are used for making the time-distance plot in Figure~\ref{fig:slice}.  
The green line in panel (e) marks the central PIL for reference. 
The PIL is directly extracted at the contour line of $B_z=0$ drawn on the photospheric $B_z$ magnetogam. The images are not shown in regular cadence since the polarities are not emerging in a regular pace. An online animation is also available. 
}\label{fig:bz_evo}
\end{center}
\end{figure*}

NOAA AR 12673 hosts a multi-polar configuration formed by different groups of flux concentrations 
emerging at different times. Five bipoles are involved in forming the central PIL.  
They start to emerge in succession from around 18:00 UT on 2017 September 3, 
surrounded by other already emerged polarities (Figure~\ref{fig:bz_evo}). 
The conjugate polarities of each bipole are determined by the criteria proposed in~\citet{Chintzoglou_2018}: 
they emerge at the same time and move away from each other. 
The earliest emerged negative polarity (N1 in Figure~\ref{fig:bz_evo}(a)) is adjacent to a pre-existing positive polarity patch 
(P0 in Figure~\ref{fig:bz_evo}(a)), while its conjugate positive polarity (P1 in Figure~\ref{fig:bz_evo}(a)) moves southwestward. 
Later on, two other bipoles, N2 P2 and N3 P3 emerge successively at the north of N1 (Figure~\ref{fig:bz_evo}(b)-(c)). 
The separation of their conjugate polarities are not as clear as that of the bipole N1 P1, 
likely because that the surrounding pre-existing polarities inhibit their motions. 
Nevertheless, one can still see the overall northeastward motion for N2 and N3, and southward motion at least for P2. 
The latter results in collision to N1. 

After the data gap caused by the solar eclipse from around 06:00 UT to 08:30 UT on 2017 September 4, another two bipoles appear on the magnetograms (see Figure~\ref{fig:bz_evo}(d)-(f)). 
The bipole N4 P4 emerges at the south of P1, 
with a clear separation identified. 
The polarity N4 seems to have collided with part of P1. The other bipole N5 P5 emerges at the northwest of N2 and N3. 
The clear separation between N5 and P5 is also observed, 
with P5 moving southeastward and mixing with pre-existing P2 and P3 which have collided with N1. 
The negative polarities N1, N2 and N3 merge gradually later.  

The above results suggest that the central PIL (green line in Figure~\ref{fig:bz_evo}(e)) 
is formed by the interaction between different bipoles, 
at least including the collision between non-conjugated polarities at two locations, one between N1 and the mixture of P2, P3 and P5, and the other one between N4 and P1. The final PIL could be classified as a collisional PIL as defined in~\cite{Chintzoglou_2018}, rather than a self-PIL within a single bipole. 
Note that we also identify collision between two nearby, pre-existing bipoles (NA PA and NB PB in Figure~\ref{fig:bz_evo}(a)-(b)), 
where clear cancellation occurs 
(see the animation associated with Figure~\ref{fig:bz_evo}) and a strongly sheared (or weakly twisted) structure forms and evolves independently from the MFR formed at the central PIL (see Section~\ref{subsubsec:sigmoid}). For simplicity, we focus on the MFR formation above the complex central PIL we describe above, and pay no much attention to the structure which is formed by the flux cancellation clearly.

\subsubsection{Appearance of a Sigmoid}\label{subsubsec:sigmoid} 

\begin{figure*}
\begin{center}
\epsscale{1.1}
\plotone{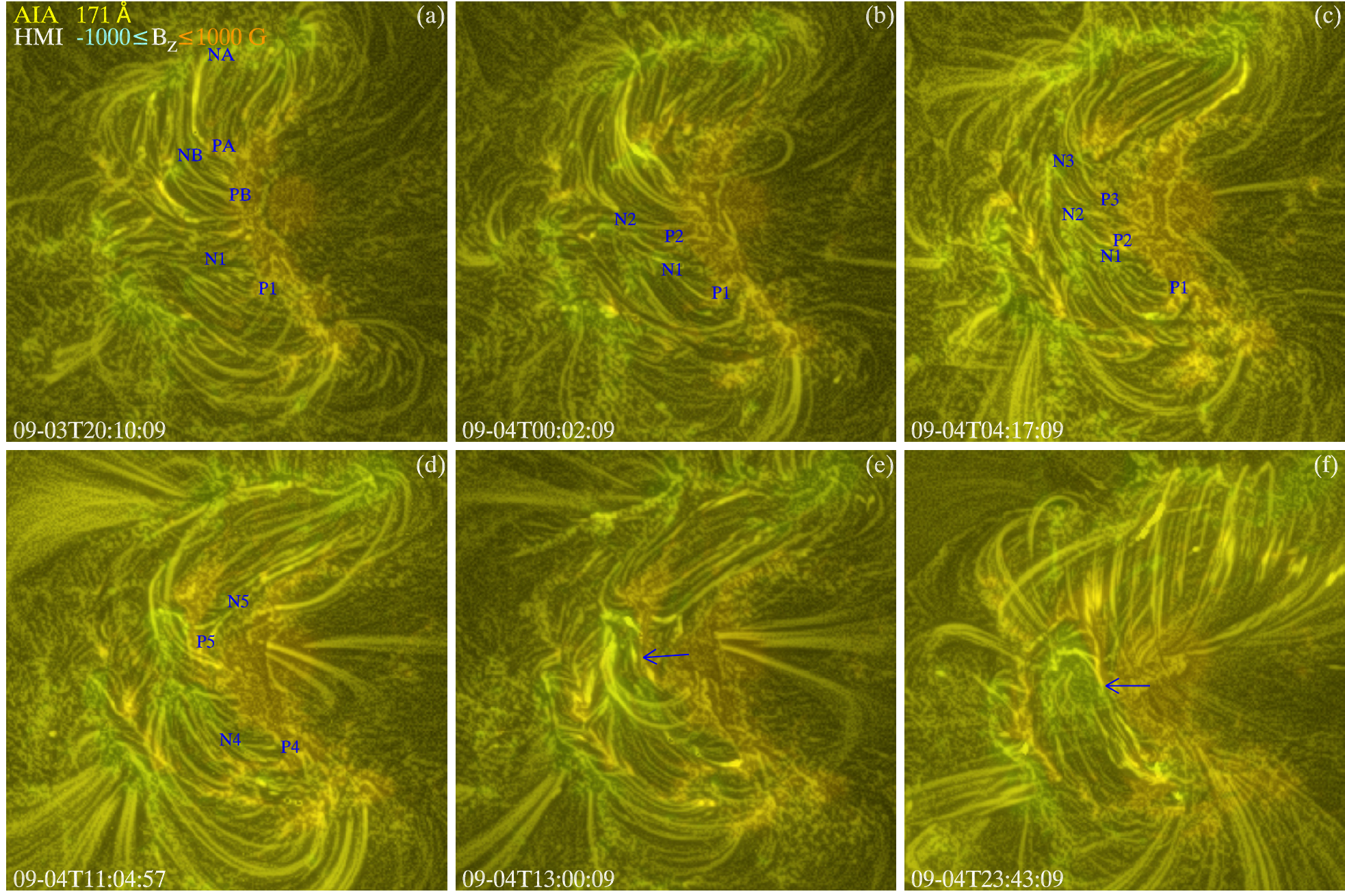}
\caption{The AIA 171 \AA~images showing the evolution of the AR. 
The images are processed by a Multi-scale Gaussian Normalization method~\citep[MGN method,][]{Morgan_2014a} to enhance the ratio of signal to noise.  
The backgrounds represent the HMI $B_z$ magnetograms. 
The blue labels in panels (a)-(d) mark the interested polarities as same as in Figure~\ref{fig:bz_evo}. 
The arrow in panel (e) indicates the brightenings along the central PIL, while the one in panel (f) marks a possible sigmoidal structure. The images are not shown in regular cadence since the polarities are not emerging in a regular pace. An online animation is also available.  
}\label{fig:euv_conn}
\end{center}
\end{figure*}

We then examine the AIA images at all EUV passbands to investigate the coronal evolution of the region.  
The connectivity of the polarities that we infer from the magnetograms is confirmed at the 171 \AA~passband (Figure~\ref{fig:euv_conn}). 
For each conjugate polarities pair, 
nearly-potential loops connecting them appear when they start to emerge (Figure~\ref{fig:euv_conn}(a)-(d)). As the flux emergence progresses, 
the loops seemingly get sheared gradually, 
and blurred by nearby dynamically evolving loops. 
Along the central PIL, brightenings occur occasionally (Figure~\ref{fig:euv_conn}(e)); 
a possible sigmoidal structure seemingly 
appears at around 09-04T23:43 UT (Figure~\ref{fig:euv_conn}(f)).

\begin{figure*}
\begin{center}
\epsscale{1.1}
\plotone{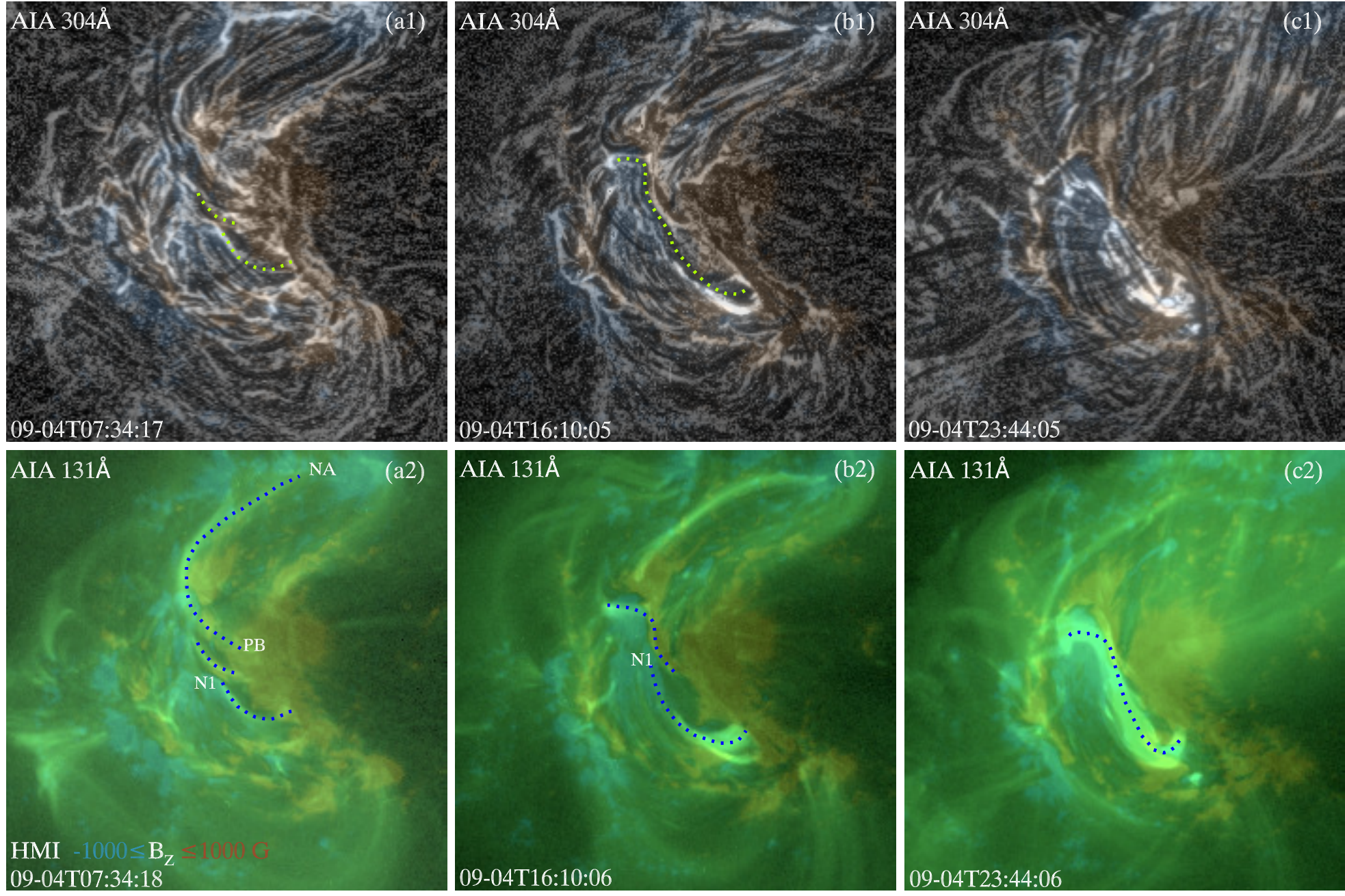}
\caption{The AIA 304 \AA~ and 131 \AA~images with the HMI $B_z$ magnetograms shown as the backgrounds. 
The AIA 304~\AA~images are processed by the MGN method. 
The dotted curves in panels (a1), (a2) and (b2) mark the sheared loops across 
the central PIL. The ones in panels (b1) and (c2) indicate a sigmoidal structure. 
An online animation is also available.}\label{fig:euv_sig}
\end{center}
\end{figure*}

At the 304 \AA~and 131 \AA~passbands, similar evolution 
is observed, although the connectivity of the interested polarities is not as clear as that in the 171 \AA~passband. 
At around 07:34 UT on 2017 September 4, two sets of sheared loops connecting N1 P1 and N2 P2 are identified in both passbands (Figure~\ref{fig:euv_sig}(a1)-(a2)). 
Another set of strongly sheared loops connecting pre-existing, non-conjugated polarities NA and PB are also visible in the 131 \AA~(Figure~\ref{fig:euv_sig}(a2)), which results from the flux cancellation between PA and NB (see Section~\ref{subsubsec:emergence}). 
At around 16:10 UT, the sheared filamentary loops at the 304~\AA~passband seem to merge near N1, 
that an elongated filament appears (Figure~\ref{fig:euv_sig}(b1)).    
Meanwhile, two sets of disconnected sheared loops are still seen in the 131 \AA~passband, 
with one connecting N3 and the mixed positive polarities next to N1, 
the other connecting N1 and P4 (Figure~\ref{fig:euv_sig}(b2)). 
At around 23:44 UT, the filament in the 304~\AA~passband is not discernible. 
In contrast, 
a continuous sigmoidal hot channel is seen in the 131~\AA~passband (Figure~\ref{fig:euv_sig}(c1)-(c2)).

The appearance of a sigmoidal filament and a hot channel indicates that an MFR starts to appear above the PIL, at least from around 15:30 
UT to 23:44 UT on 2017 September 4 (see Figure~\ref{fig:euv_sig} associated animation). 
It may be formed through the magnetic reconnection (manifested as the merging) between different loops at the collision locations. 
The process may occur in the relatively cold, lower solar atmosphere as the MFR 
firstly appears in the 304 \AA~passband which images the chromosphere~\citep{Lemen_etal_2012}. 
It then may rise and get heated to appear in the high-temperature wavelength such as 131 \AA. 
In the next section, we resort to the NLFFFs modeling to quantitatively study the details of the MFR formation.

\subsection{Mechanism of the MFR Formation}\label{subsec:mechnism}

\subsubsection{Formation details revealed by 3D NLFFFs}\label{subsubsec:NLFFF} 

We extrapolate a time series of 3D NLFFFs, 
based on which the observed connectivity is well reproduced (Figure~\ref{fig:bf_mfr}).  
When the bipoles start to emerge, the magnetic field lines connecting the conjugated polarities are nearly potential as evidenced by, 
e.g., the small $T_w$ (of around $-0.2$) that the field lines from N1 have (Figure~\ref{fig:bf_mfr}(a1)). 
With P2 and P3 moving southward, field lines connecting N2, P2, N1 and P1 appear, 
showing a U-shaped bottom near N1 and $T_w$ of around $-0.4$ 
(Figure~\ref{fig:bf_mfr}(b1)-(c1)). 
This confirms that a merging process occurs at the collision location (between N1 and P2).  
Similar process occurs at the other collision location (between N4 and P1), 
as the field lines passing through N2, P2, N1, P1, N4 and P4 appear later, 
presenting a ``sea-serpent'' configuration with U-shaped bottoms near N1 and N4. 
Their twist number increases to around $-0.8$ (Figure~\ref{fig:bf_mfr}(d1)). 
Note that the small positive polarity next to N1, labeled as P0, is found to connect to the remote pre-existing negative polarity that is not involved into the MFR formation later.

At 09-04T09:46 UT, the field lines connecting N3, P3, N1, P1, N4 and P4 appear along the PIL, 
showing $|T_w| \geqslant 1$ (Figure~\ref{fig:fr_cmp}(a)), 
which indicates a further merging between N1 and P3 and the formation of an MFR seed. 
The MFR does not have a clear, closed QSL (high Q lines) boundary (Figure~\ref{fig:fr_cmp}(a1) and (a2)),  
indicating that its formation is on-going. 
The bottom of the QSL touches the photosphere, 
revealing the presence of a BP configuration at the collision location 
(between N1 and the mixed positive polarities). 
As time goes on, 
the cross section of the MFR (the region with $|T_w| \geqslant 1$) keeps growing bigger, with the high Q boundary getting clearer and closed gradually (Figure~\ref{fig:fr_cmp}(b)-(f)). 
The cross section shows inhomogeneous $T_w$ distribution, with a region owning stronger $T_w$ than its surrounding. 
Note that the connectivity of the later emerged bipole N5 P5 is also reproduced (see Figure~\ref{fig:fr_cmp}(b)). Although P5 moves southward and gets mixed with P2 and P3, no MFR field line connecting to 
N5 is found, indicating that the bipole does not contribute magnetic flux to the MFR directly.

We further calculate the axial flux $\Phi$ and average twist number $\overline{T}_w$ at several selected moments. 
To calculate the parameters, one need to determine the boundary of the MFR. 
For a full-fledged MFR, 
the clear high Q lines enclosing it can be taken as its boundary. 
Whilst for a MFR during formation, it may not have closed high Q boundary (see Figure~\ref{fig:fr_cmp}). 
Thus, we determine the MFR boundary here through combining the existing high Q lines and the contour of the threshold of $T_w$ manually 
(see also~\cite{Liu_2018d}). The threshold value of $T_w$ is set to vary from $-1.2$ to $-1$ to ensure the contour line and the unclosed high Q line connect smoothly. 
At each moment, we repeat the boundary identification and the calculation at three vertical slices, 
and get three values for each parameter. 
The mean value of them is taken as the final value and the standard error is taken as part of the error. 
The rest of the error is considered from two sources, 
one is the uncertainty of manual boundary identification, 
the other is the photospheric noise and the orbital variations that propagate through the NLFFFs model. 
The former is estimated to be $10\%$ and $3\%$ for $\Phi$ and $\overline{T}_w$ according to the results in~\cite{Liu_2018d}.  
For the latter,~\citet{Wiegelmann_2010b} reveals that a uniform noise of around $5\%$ of the maximum transverse field introduces about $4\%$ error to the extrapolated field. 
Taking this result, we get an error of $4\%$ to $\Phi$, 
and a negligible small error to $\overline{T}_w$.

Figure~\ref{fig:twist_flux} shows the final results. 
The weakly sheared structure during the very early emergence phase 
(e.g., at around 09-04T04:10) has $\overline{T}_w$ and $\Phi$ 
as small as $-0.4$ and $1.1\times10^{18}$~Mx, respective. 
After the data gap, 
the two parameters start to increase significantly. 
Until 12:58 UT, $\Phi$ displays a gradual increase to around $0.6\times10^{20}$~Mx with an average growth rate around $0.06\times10^{20}$~Mx 
h$^{-1}$, while $\overline{T}_w$ rises to $-1.5$ rapidly with an average growth rate around $-0.14$ 
h$^{-1}$. From 12:58 UT to 18:34 UT, $\Phi$ keeps increasing to $2.1\times10^{20}$~Mx with a faster growth rate around $0.3\times10^{20}$~Mx 
h$^{-1}$, while $\overline{T}_w$ remains steady around $-1.5$. Later on, both $\Phi$ and $\overline{T}_w$ remain relatively steady. 
We conclude that a well-shaped MFR has been formed by this time.

The evolution of the current density is also checked (Figure~\ref{fig:jv}). 
It is found that the current firstly accumulates near the collision location 
(between N1 and the nearby mixed positive polarities) (Figure~\ref{fig:jv}(a)-(c)), 
then grows along the whole PIL gradually along with the MFR formation (Figure~\ref{fig:jv}(d)-(f)). 
The strong current region at around 09-04T22:34 UT resembles the shape of the MFR (Figure~\ref{fig:jv}(f)), 
indicating that the volume current along the MFR has been built up.

The above NLFFFs result reveals a similar process as indicated in the EUV images, i.e.,  
the MFR is highly possible to be formed through magnetic reconnection at the collision locations, 
probably with a shearing process accompanied.   
It further reveals that the MFR seed is formed by around 09-04T09:46 UT. 
Subsequently, its magnetic flux and twist number keep growing with time. 
Finally, the MFR  has relative steady $\Phi$ and $\overline{T}_w$ of around $2.1\times10^{20}$~Mx and $-1.5$, respective.

\subsubsection{Topology of the MFR}\label{subsubsec:topo}

It is shown that along with the MFR formation (Figure~\ref{fig:fr_cmp}(a2)-(f2)), 
a BP configuration exists at one collision location (between N1 and its non-conjugated positive polarities),  
at which the discrete sheared loops get merged. 
The other collision location, where merging also occurs, is the interface of N4 and P1. 
This indicates that the MFR probably owns a complex configuration with more than one BP. 
We select a moment (2017-09-04T16:10 UT) to analyze the topology of the MFR in detail. 
We do identify two elongated BPs at the central PIL on the photosphere  
(BP1 and BP2 in Figure~\ref{fig:bp_cmp_bh}(a)) by the BP criterion, $\mathbf{B}_h \cdot \mathbf{\nabla}_hB_z|_{PIL}>0$ \citep{Titov_1993, Titov_1999}, 
one between N1 and the mixed positive polarities (BP1), 
the other between N4 and P1 (BP2), consistent with the locations where the collision and merging occur.  
The horizontal field at the BPs clearly show inverse configuration (Figure~\ref{fig:bp_cmp_bh}(b)).  
Note that the MFR in fact touches the bottom boundary of the NLFFFs, which is not exactly the same as the photosphere, 
so that we compare the BPs on the two layers. It is found that the BPs on the former have almost the same patterns as the ones on the latter,  
with a coincidence ratio up to $89\%$ (Figure~\ref{fig:bp_cmp_bh}(c)).  
This suggests that the topological signatures on the photosphere are kept by the extrapolated NLFFF bottoms to a large extent.

Detailed topology of the MFR is displayed in Figure~\ref{fig:mfr_topo}.  
Three sets of representative field lines are shown (Figure~\ref{fig:mfr_topo}(a)-(b)). Two sets of them 
pass through BP1, having $T_w$ of $-1.9$ and $-1.4$ (indicated by pink and blue arrows in Figure~\ref{fig:mfr_topo}), respective. The other set passing through BP2 has $T_w$ of $-1$ (indicated by orange arrow in Figure~\ref{fig:mfr_topo}). 
The inhomogeneous $T_w$ distributions in the cross section of the MFR (Figure~\ref{fig:fr_cmp}) could be due to the fact that the helical field lines are merged from different sets of discrete loops which have different degrees of shear. 
Meanwhile, sheared field lines (in silver color in Figure~\ref{fig:mfr_topo}(b)) also exist near the BPs, 
having orientations similar to the nearby twisted field lines. 
The 3D distribution of Q (Figure~\ref{fig:mfr_topo}(c)) shows larger values in the regions extending from the BPs 
to the lower atmosphere, indicating 
extending BP QSLs. 
Except the two BPs, 
a very low-lying X point (with a height around 0.5~Mm), where two QSLs intersect, is found between them {(Figure~\ref{fig:mfr_topo}(d)-(f))}, 
suggesting an HFT configuration. 
Both BPs and HFT with high Q values are the most likely locations where magnetic reconnection takes place~\citep[e.g.,][]{Titov_1993,Fan_2004}. 
While the reconnection at the photospheric BPs should result in flux cancellation, 
that at the low-lying HFT, 
the height of which is comparable to the critical height for the submergence of reconnected short loops~\citep{VanBallegooijen_1989}, 
may also result in flux cancellation.  

In short, 
we analyzed the detailed topology of the MFR. 
The MFR owns two BPs formed at the collision locations and an HFT between them, all favoring the reconnection. 
It further supports that the reconnection 
happening at the BPs, may also at the HFT,  
leads to the formation of the MFR. 

It should be noted that the above results are derived based on the 
model that has the non-linear force free assumption, 
which generally is not satisfied in the lower atmosphere. 
To check the reliability of the results, 
we compare the extrapolated MFR and the MFR proxy imaged by the EUV passbands (see Figure~\ref{fig:mfr_euv}). 
It is found that the overall shape of the filament in the 304~\AA~passband and the hot channel in the 131~\AA~passband is reproduced by the extrapolated MFR to a large extent. Furthermore, 
the force-free and divergence-free conditions for all extrapolations we performed are also checked using the criteria proposed in~\citealp{Wheatland_2000}~\citep[see examples in][]{Lliu_2017}, and are found to be small. 
We conclude that the quality of the NLFFFs extrapolations here is acceptable.  

\subsubsection{Shearing and Cancellation Resulted from the Collision}\label{subsubsec:motions} 

\begin{figure*}
\begin{center}
\epsscale{1.1}
\plotone{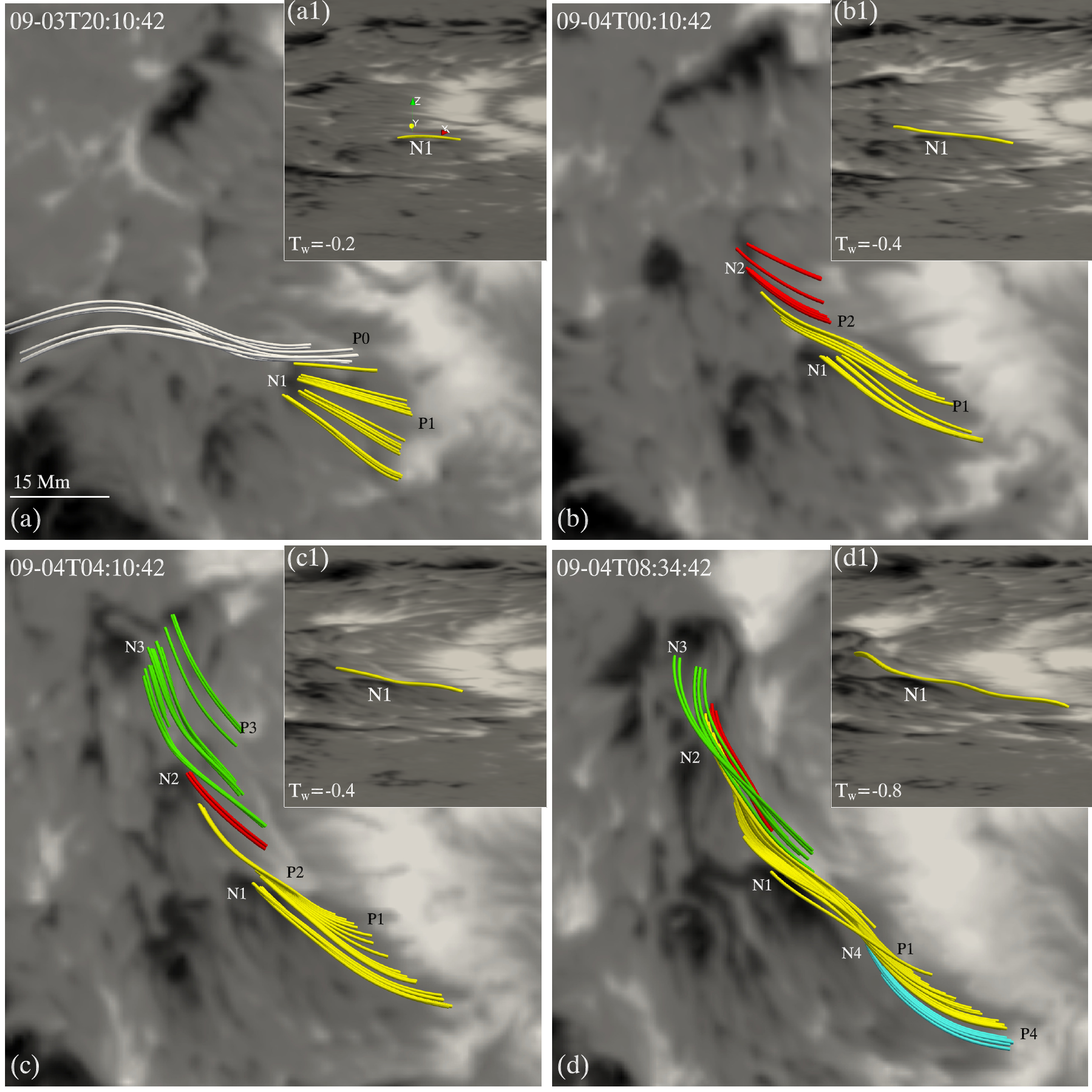}
\caption{Representative NLFFFs lines traced from the emerging bipoles, 
with different colors representing field lines from different bipoles. 
The backgrounds are $B_z$ on the photosphere,  
saturating at $\pm 2000$~Gauss with black (white) patches for the negative (positive) polarities. 
The insets (a1)-(d1) at the upper-right corners display representative field lines passing through 
N1 in another view angle.}\label{fig:bf_mfr}
\end{center}
\end{figure*}

\begin{figure*}
\begin{center}
\epsscale{1.1}
\plotone{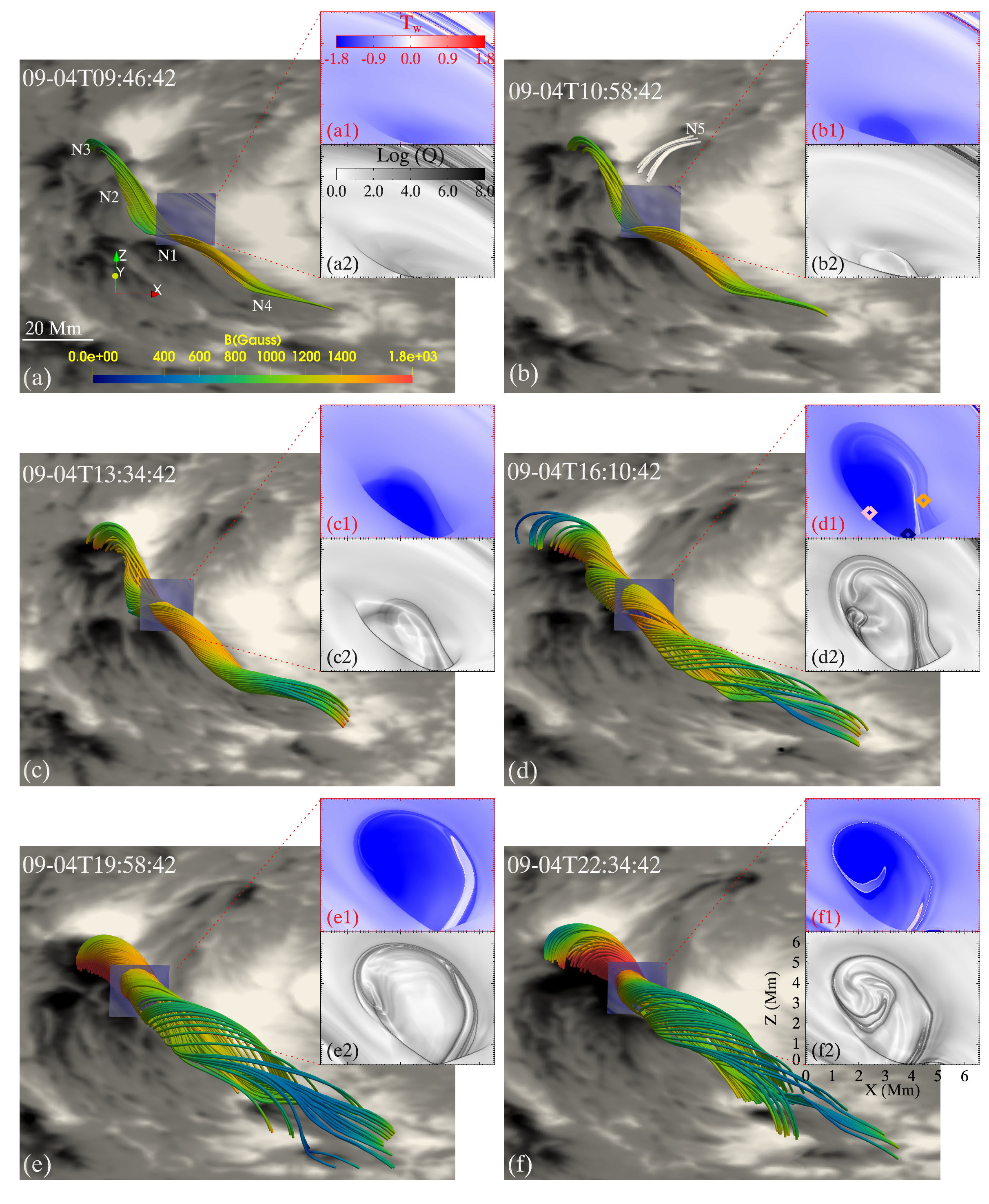}
\caption{The evolution of the MFR above the emerging region. Representative field lines constituting the MFR are plotted in iridescence, varying with the field strength. The upper-right insets (a1)-(f1) and (a2)-(f2) 
display distributions of $T_w$ and $Q$ in vertical cuts across the interface of N1 and the mixed positive polarities. 
Colored diamonds in panel (d1) mark the positions where the field lines in Figure~\ref{fig:mfr_topo} thread. 
The silver lines in panel (b) are the representative field lines from the bipole N5 P5 which emerges after the data gap.}\label{fig:fr_cmp}
\end{center}
\end{figure*}

\begin{figure*}
\begin{center}
\epsscale{1.1}
\plotone{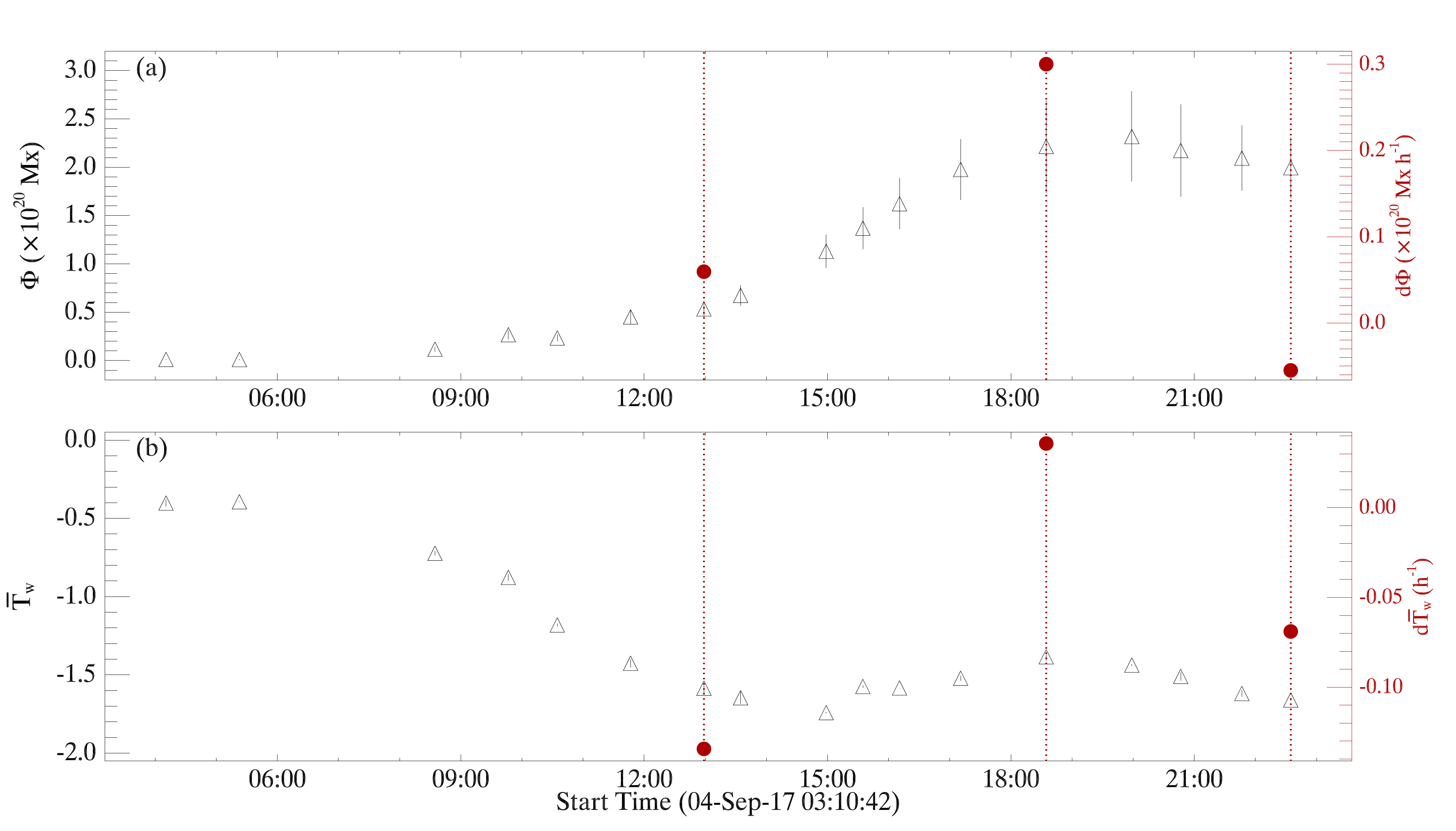}
\caption{Evolution of the MFR parameters. (a): the axial flux $\Phi$ of the MFR. (b): the average twist number $\overline{T}_w$ of the MFR. 
The red vertical lines divide the time into three periods. 
The corresponding red dots give the average change rate during each period.}\label{fig:twist_flux}
\end{center}
\end{figure*}

\begin{figure*}
\begin{center}
\epsscale{1.2}
\plotone{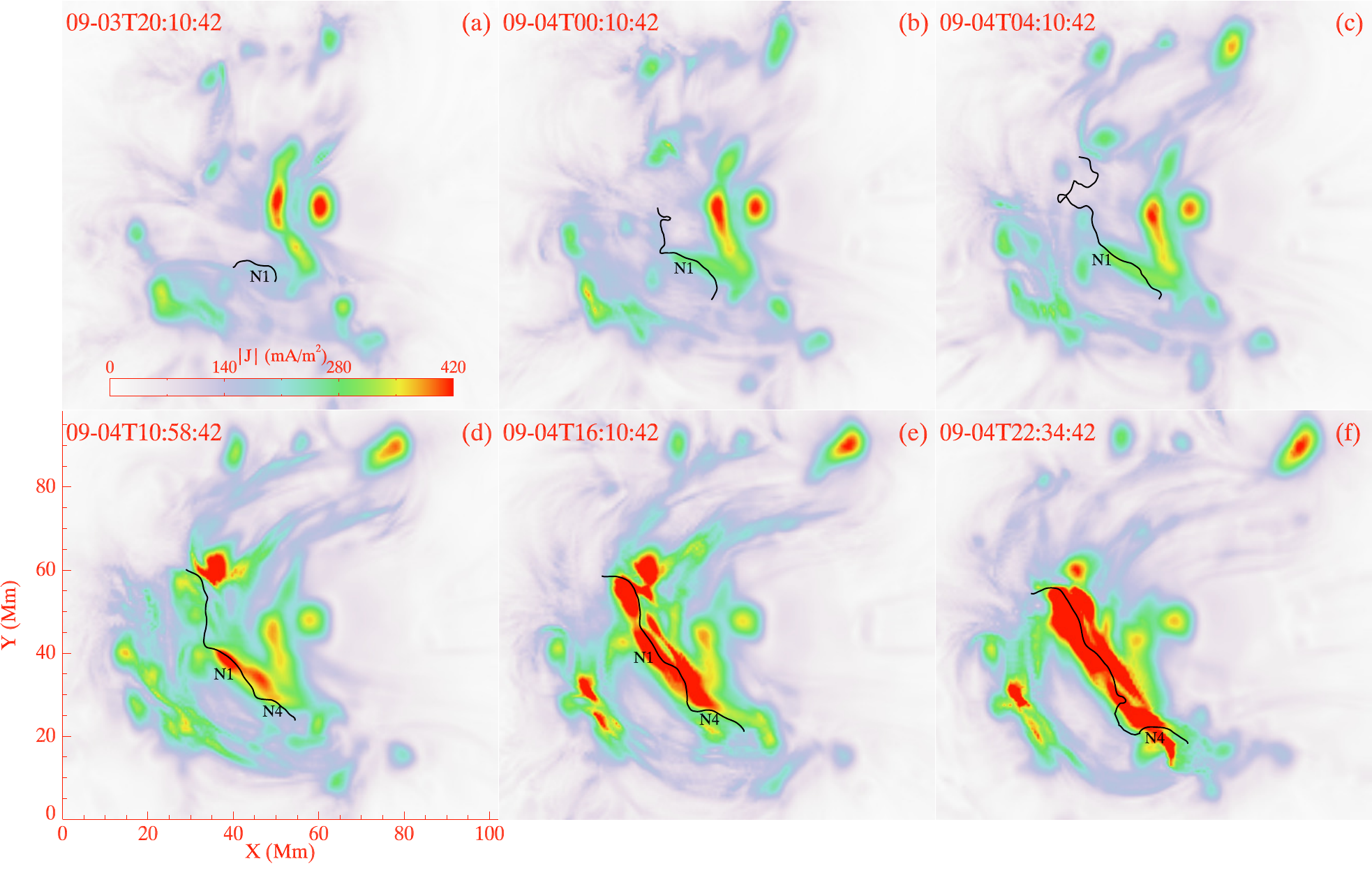}
\caption{Evolution of the current density $J$. The value displayed in each pixel is vertically integrated from the bottom boundary to the pixel at 10~Mm in the NLFFFs using $\displaystyle \mathbf{J}=\frac{1}{\mu_0} \Sigma_h \nabla\times\mathbf{B}(h)$ since the MFR is well below 10~Mm. The black line in each panel outlines the PIL, which is directly extracted at the contour line of $B_z=0$ drawn on the photospheric $B_z$ magnetogam.}\label{fig:jv}  
\end{center}
\end{figure*}

\begin{figure*}
\begin{center}
\epsscale{1.1}
\plotone{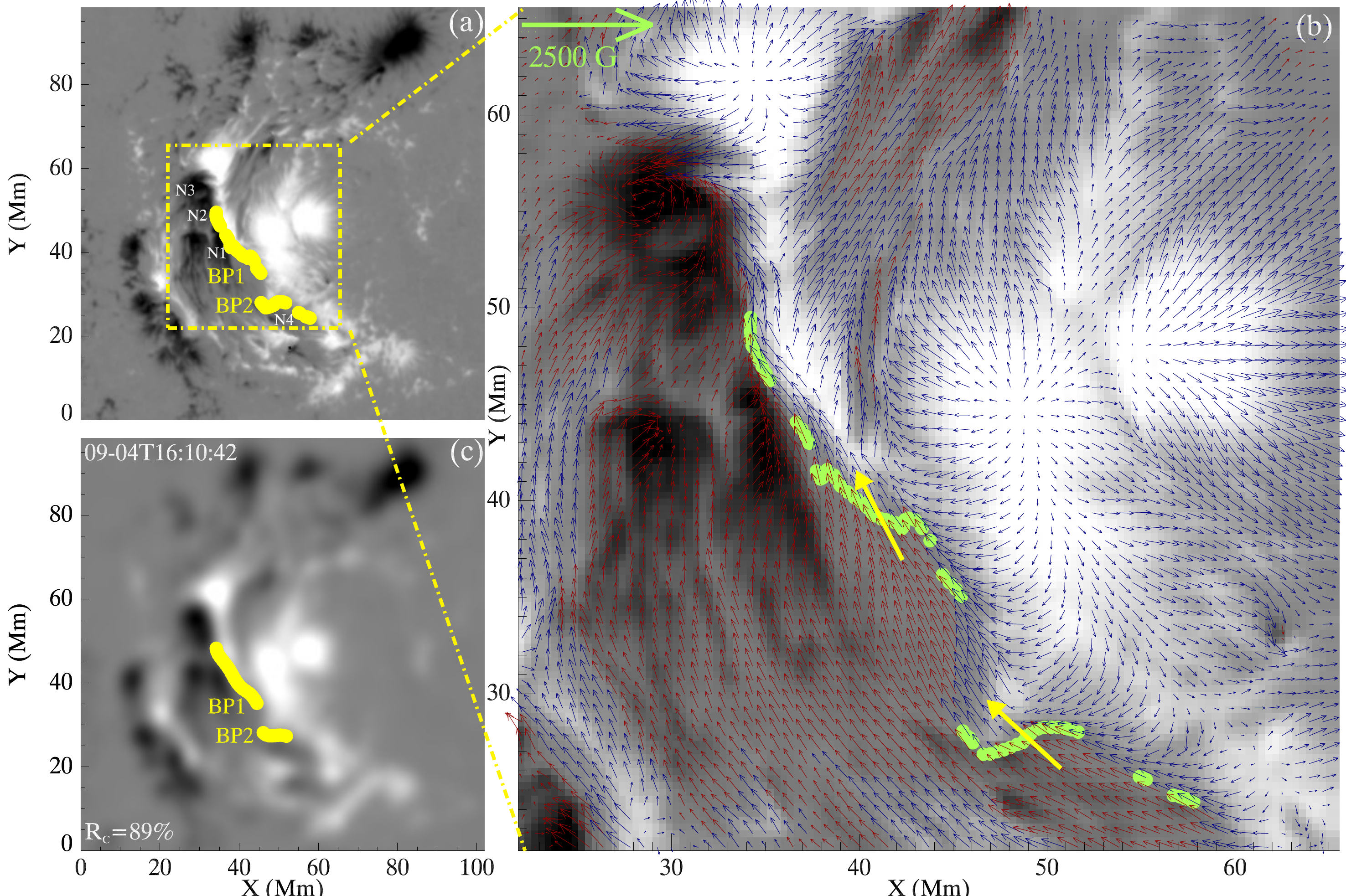}
\caption{The BPs identified on the central PIL at 2017-09-04T16:10 UT. (a) The BPs identified on the photoshpere. (b) The vector magnetic field in the core region on the photosphere. (c): The BPs identified on the bottom boundary of the NLFFFs. $B_z$ are plotted as the backgrounds of these panels, saturating at $\pm 2000$ Gauss. Two elongated BPs, BP1 and BP2 are marked as thick yellow lines in panels (a) and (c), and green lines 
in panel (b). The blue (red) arrows in panel (b) are the horizontal component ($B_h$) of the magnetic field coming from (going into) the positive (negative) polarity patches. The yellow thick arrows in panel (b) point out the orientations of 
$B_h$ passing through the BPs. 
 }\label{fig:bp_cmp_bh}
\end{center}
\end{figure*}

\begin{figure*}
\begin{center}
\epsscale{1.1}
\plotone{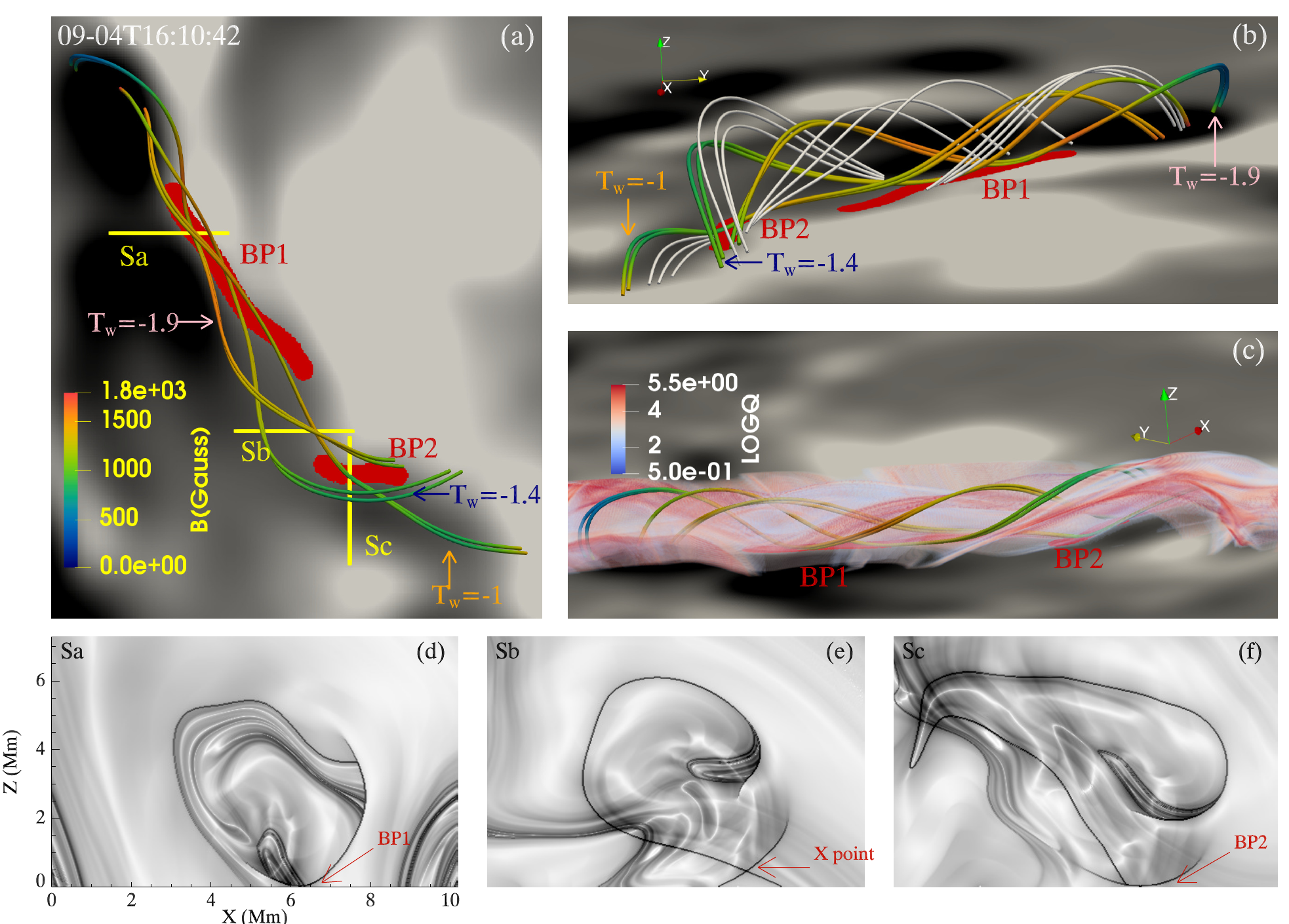}
\caption{The topology of the MFR at 2017-09-04T16:10 UT. (a): representative field lines passing through the BPs. Yellow lines Sa, Sb and Sc mark the positions of three vertical cuts used in panels (d), (e) and (f). The red thick lines mark the two BPs. (b): another view of the field lines shown in panel (a). Representative field lines of the sheared arcades near the BPs are also shown in silver. (c): 3D Q in a cylindrical region containing the MFR. 
(d) - (f): distributions of Q in the three vertical cuts. Twist numbers of the field lines in panels (a) and (b) are marked in the same colors as the corresponding diamond symbols in Figure~\ref{fig:fr_cmp}(d1). 
}\label{fig:mfr_topo} 
\end{center}
\end{figure*}

\begin{figure*}
\begin{center}
\epsscale{1.1}
\plotone{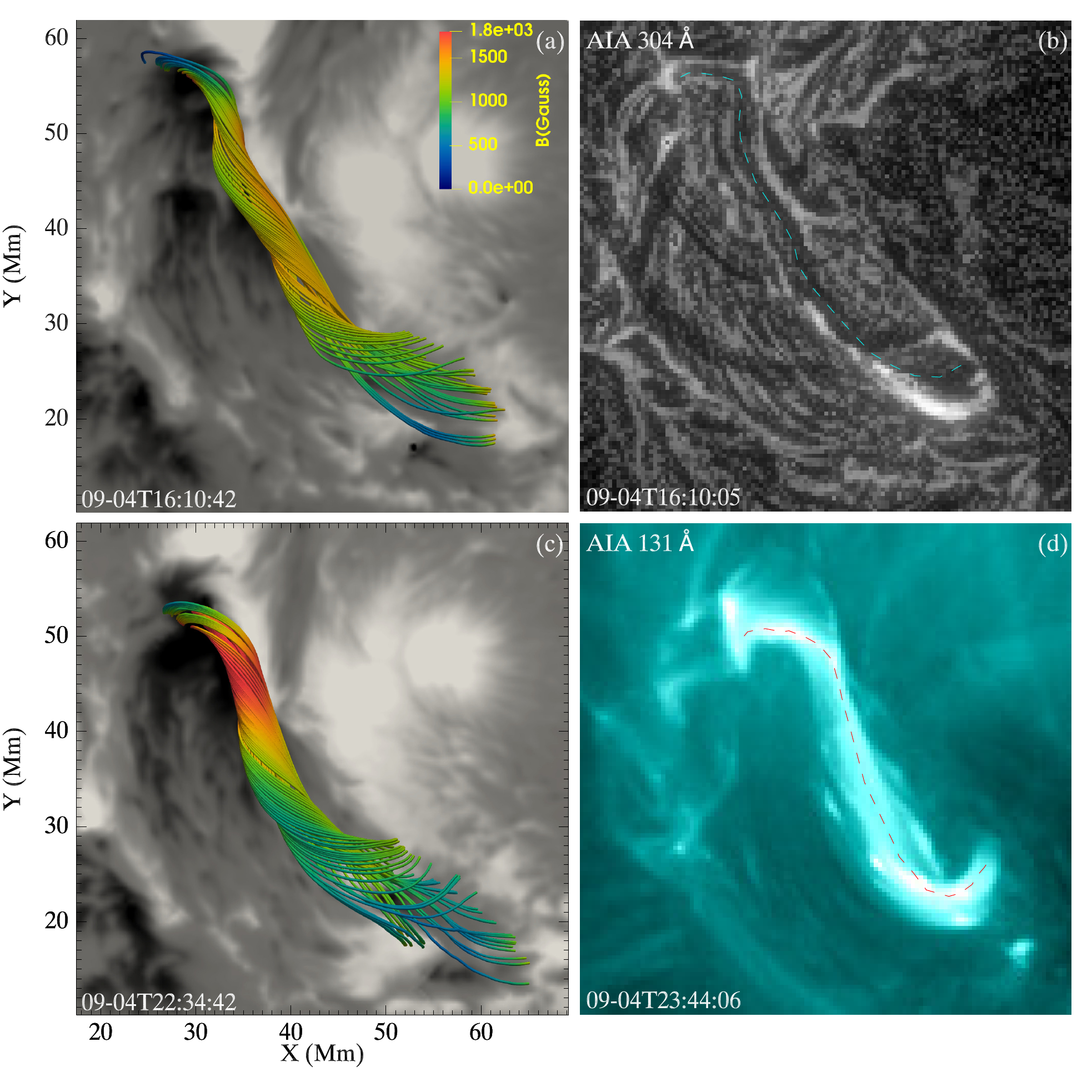}
\caption{The upper row: the extrapolated MFR 
and a filament observed in the 304~\AA~passband 
at around 09-04T16:10 UT. The 304~\AA~image 
is processed by the MGN method. 
The lower row: 
the extrapolated MFR at around 09-04T22:34 UT and a hot channel observed in the 131~\AA~passband at around 09-04T23:34 UT. 
The hot channel in panel (d) seems to be 
ignited in some heating process, during which the non-linear force-free assumption might not be satisfied, 
thus we show the extrapolated MFR about one hour ago (in panel (c))  
for comparison. The dashed lines in panels (b) and (d) indicate the MFR.}\label{fig:mfr_euv}
\end{center}
\end{figure*}

\begin{figure*}
\begin{center}
\epsscale{1.1}
\plotone{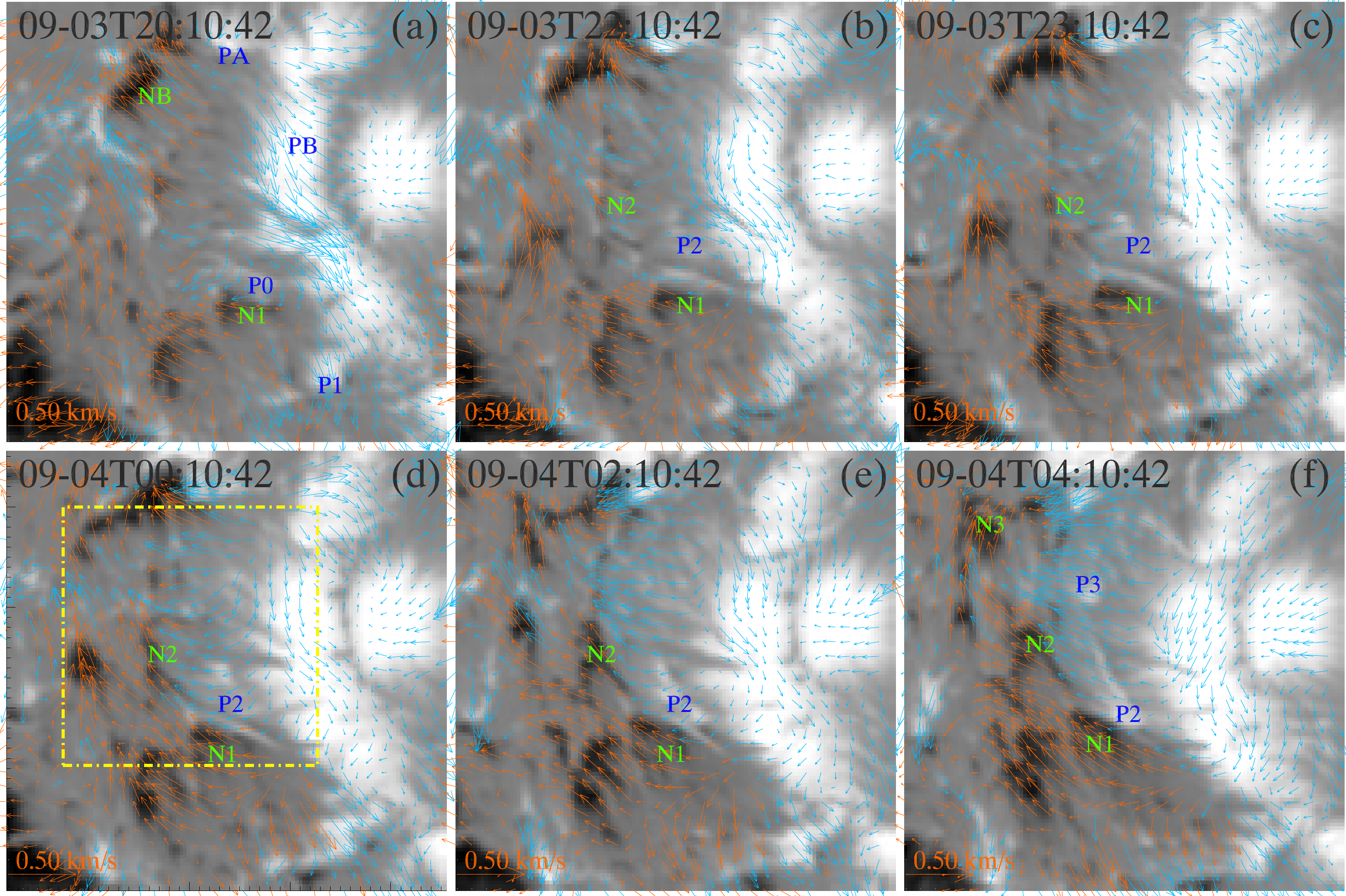}
\caption{Photospheric horizontal flow velocities calculated by DAVE4VM method before the data gap.  
The backgrounds are $B_z$, saturating at $\pm 2000$~Gauss with negative (positive) polarities shown in black (white). 
The arrows display the horizontal velocity field 
with orange (cyan) for the negative (positive) polarity. 
The yellow rectangle in panel (d) marks the FOV of Figure~\ref{fig:bz_cancel}.}\label{fig:velocity1}
\end{center}
\end{figure*}

We further explore the possible driver of the MFR formation through analyzing the photoshperic velocities calculated by DAVE4VM~\citep{Schuck_2008}. 
It is found that the separation between N1 and P1 has a velocity of around 0.3~km/s for each polarity (Figure~\ref{fig:velocity1}(a)),   
while that between N2 and P2 is slower, with the velocities of around 0.2~km/s and 0.1~km/s for N2 and P2, respective (Figure~\ref{fig:velocity1}(b)). 
After the collision between the oppositely moving N1 and P2, 
N1 starts to shear against P2, 
pushing P2 to turn toward northeast gradually (Figure~\ref{fig:velocity1}(b)-(f)). 
This is exact a ``collisional shearing" process, 
during which a BP appears at the collision location and the discrete loops from N1 P1 and N2 P2 get sheared and merge there (see NLFFFs in Figure~\ref{fig:bf_mfr}(b)). 
The separation of N3 P3 is not captured as they may be blocked by the pre-existing large 
positive polarity at the west (Figure~\ref{fig:velocity1}(f)).

\begin{figure*}
\begin{center}
\epsscale{1.1}
\plotone{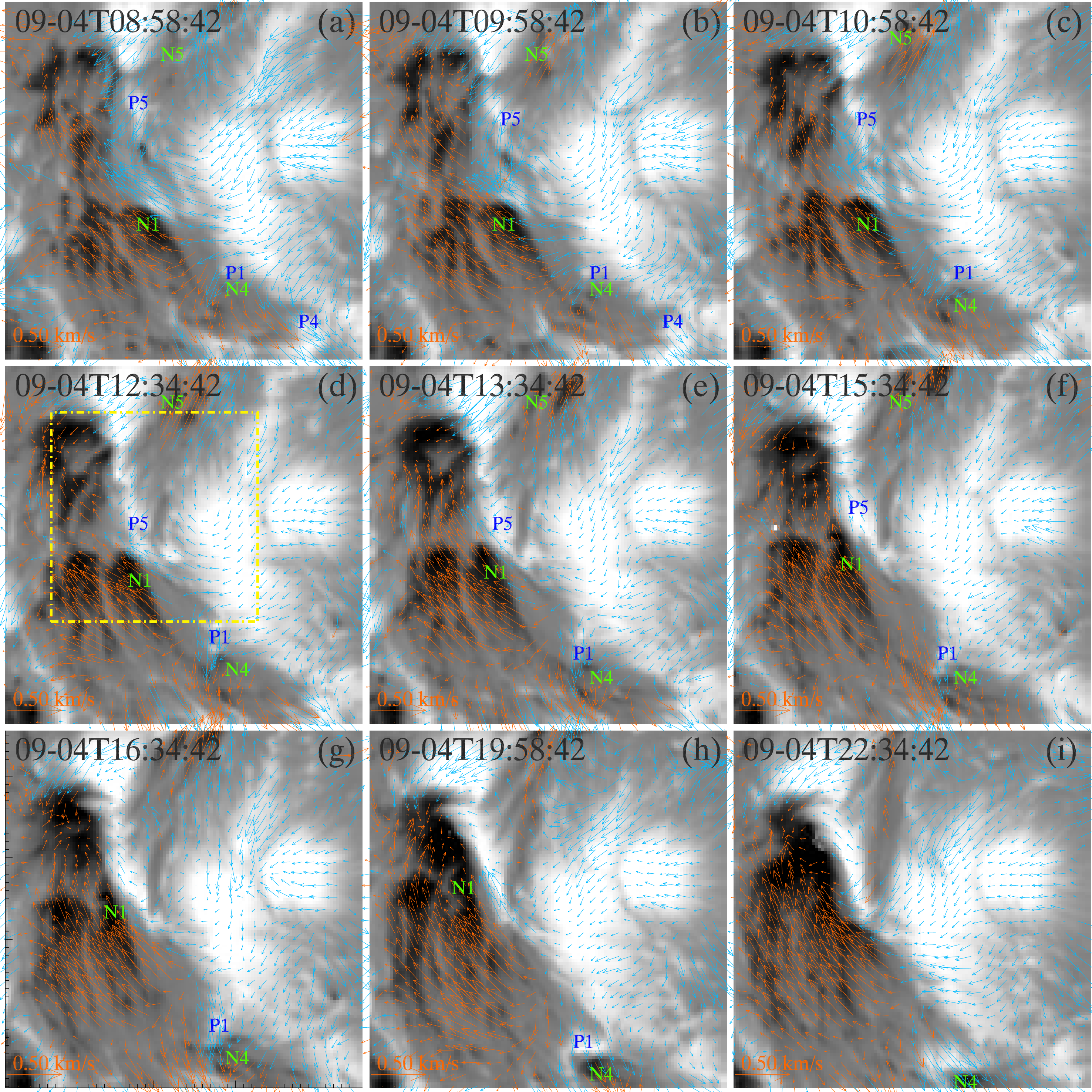}
\caption{
Photospheric horizontal flow velocities calculated by DAVE4VM method at nine selected times after the data gap. Similar layouts as Figure~\ref{fig:velocity1}. 
The yellow rectangle in panel (d) marks the FOV of Figure~\ref{fig:bz_cancel}.}\label{fig:velocity2}
\end{center}
\end{figure*}

For N4 P4 that appear after the data gap, 
while P4 departs from N4 with a velocity of around 0.4~km/s, N4 seems to have collided with P1 and gets sheared against P1, 
being pushed toward southwest with a velocity of around 0.2-0.3~km/s (Figure~\ref{fig:velocity2}(a)-(c)). 
The BP2 appears at the collision location where the field lines connecting N1 P1 and N4 P4 get merged (see Figure~\ref{fig:bf_mfr}(d)).  
The separation of N5 P5 is also captured. 
P5 firstly moves southward with a velocity of around 0.3-0.4~km/s (Figure~\ref{fig:velocity2}(a)-(c)), 
then slows down to around 0.1~km/s gradually (Figure~\ref{fig:velocity2}(d)-(e)). After reaching the mixed positive polarities next to N1, 
P5 gets sheared against them and turns toward northeast gradually (Figure~\ref{fig:velocity2}(e)-(i)). 
In general, the magnitude of both separation and shearing velocities is of around several hundreds meters per second, comparable to the results in~\citet{Chintzoglou_2018}. After the onset of collision, the separation motions gradually reduce, while the shearing motions seem to be persistent. These are also consistent with the findings in~\citet{Chintzoglou_2018}.  
The results suggest that the continuous collisional shearing near the two BPs may be the driver that adds the magnetic flux and twist into the MFR continuously.

\begin{figure*}
\begin{center}
\epsscale{0.9}
\plotone{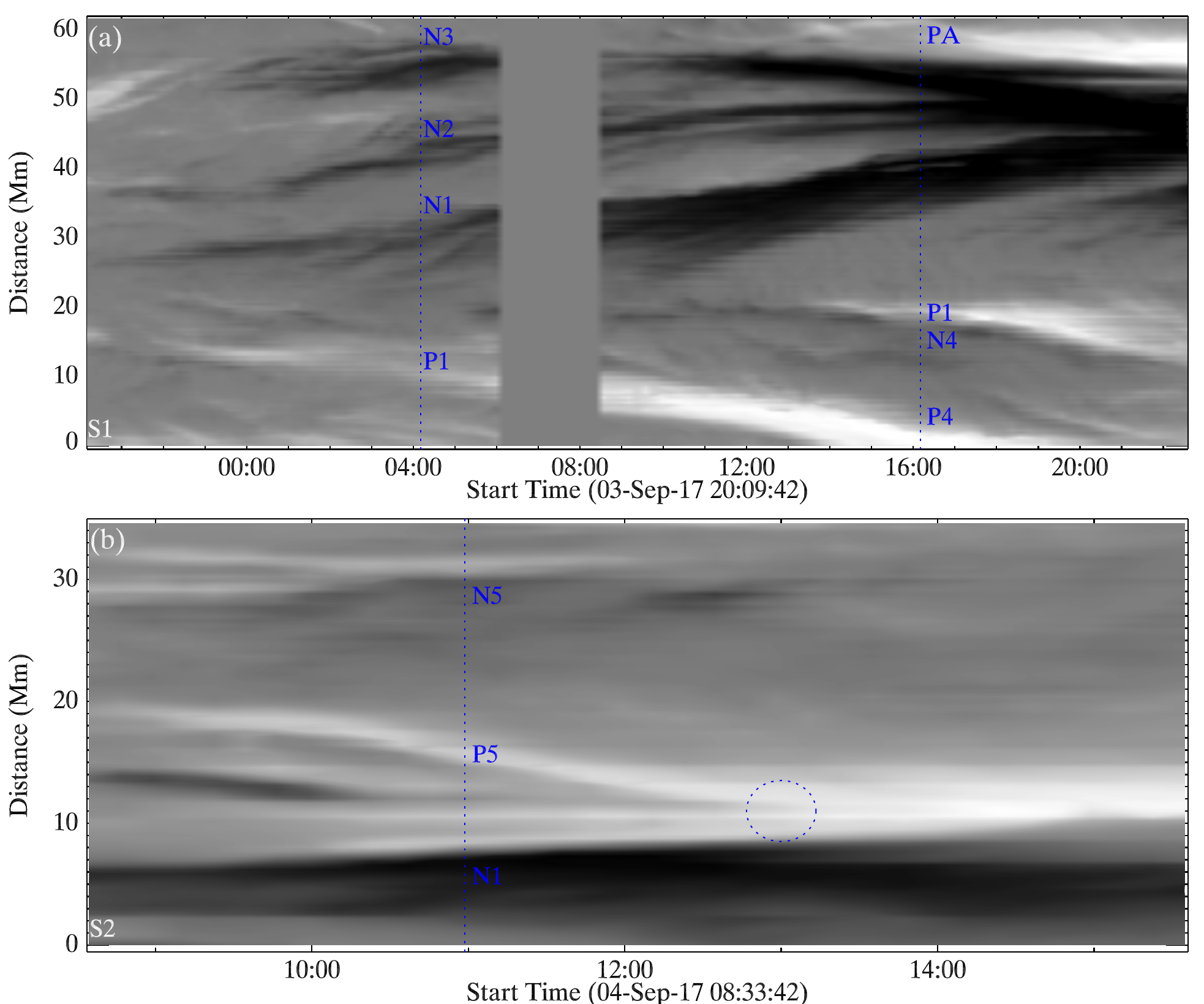}
\caption{Time-distance plots of two slits  
on the photospheric $B_z$ magnetograms, with $B_z$ saturating at $\pm 2000$ Gauss. 
The positions of the slits are shown in Figure~\ref{fig:bz_evo}. The blue eclipse in panel (b) indicates the timing when P5 collides 
with the mixed positive polarities next to N1.}\label{fig:slice} 
\end{center}
\end{figure*}

Additionally, we take two slits across different bipoles on the $B_z$ magnetograms and make their time-distance plots
(Figure~\ref{fig:slice}). 
The separation of N1 P1 and of N4 P4, and the colliding between N4 and P1 are clearly seen (Figure~\ref{fig:slice}(a)). 
The grouping of the northeastward moving N1, N2 and N3 can be attributed to the blocking of the pre-existing PA (Figure~\ref{fig:slice}(a)). 
The separation of N5 P5, and the colliding of P5 to the mixed positive polarities near BP1 are also identified (Figure~\ref{fig:slice}(b)). 
Strikingly, the colliding timing of P5 (around 09-04T13:00 UT, 
indicated by the blue eclipse in Figure~\ref{fig:slice}(b)) coincides well with the timing after which 
the MFR axial flux increases faster (see Figure~\ref{fig:twist_flux}(a)), 
suggesting that this colliding speeds up the MFR formation. 
The results unambiguously evidence that the collision drives the formation of the MFR. 

\begin{figure*}
\begin{center}
\epsscale{1.1}
\plotone{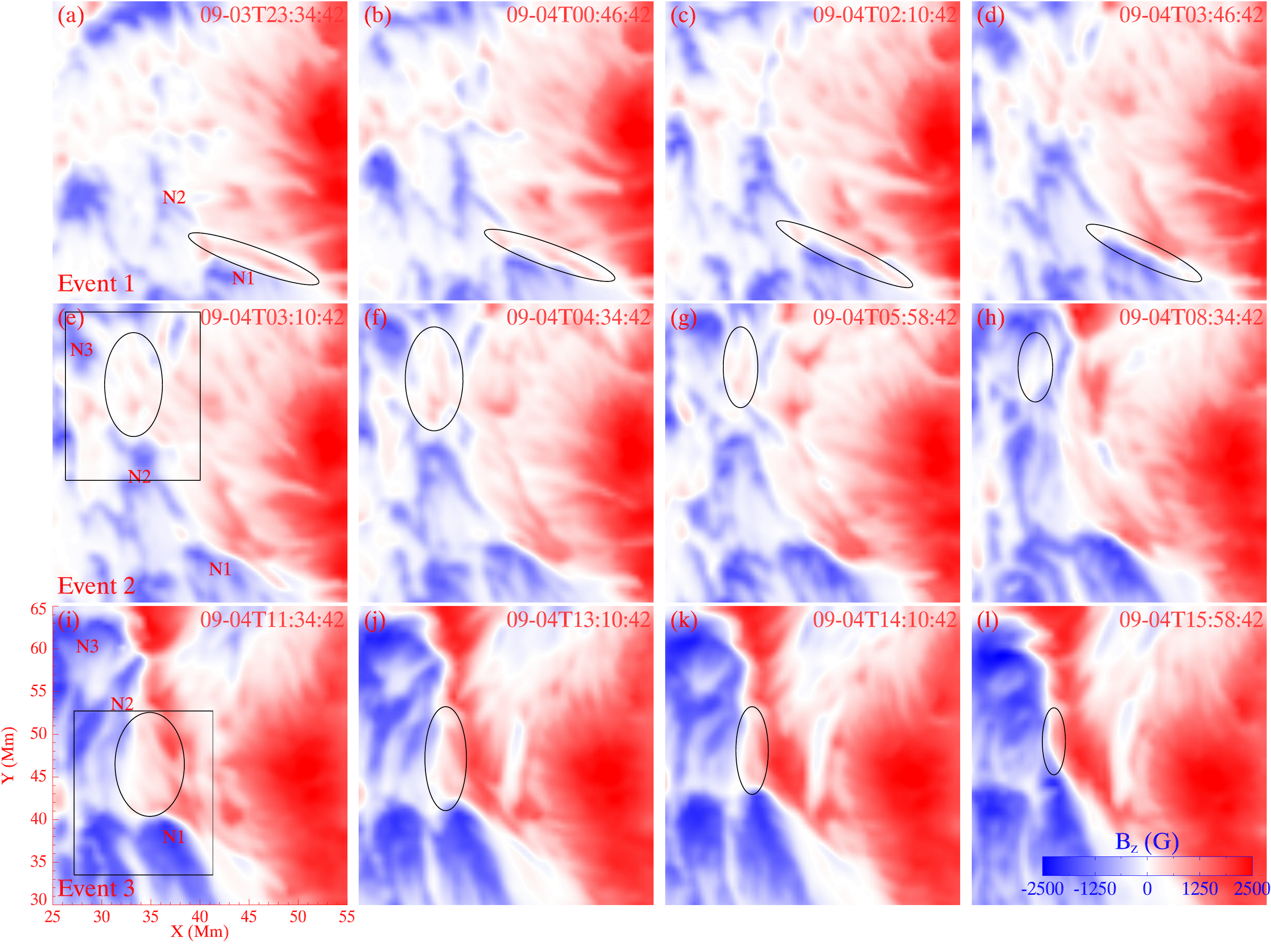}
\caption{Three identified flux cancellation events. 
Each column shows $B_z$ maps at selected moments for one event.  
The black ellipse points out 
the region where the positive polarities disappear. 
Disappearance of the corresponding negative polarities is covered by the simultaneous flux emergence thus can not be identified. 
The black rectangular in panel (e) marks the FOV of Figure~\ref{fig:bri_1700}(a)-(f). The one in panel (i) marks the FOV of Figure~\ref{fig:bri_1700}(g)-(l). An online animation is also available.}\label{fig:bz_cancel}
\end{center}
\end{figure*}

Flux cancellation is expected during the ``collisional shearing".  
After inspecting the central PIL region, 
we do identify a few cancellation events (Figure~\ref{fig:bz_cancel}), 
although cannot estimate the exact canceled amount due to the continuous flux emergence. 
The first event occurs at the very early emergence phase, during which an elongated, 
spindly positive flux patch in P2 disappears gradually 
(Figure~\ref{fig:bz_cancel}(a)-(d)). 
The corresponding canceled flux in N1 is covered by simultaneous emerging flux. 
This event corresponds well with the ``collisional shearing" between N1 and P2, and the resulting merging that creates serpentine-shaped loops there (see Figure~\ref{fig:bf_mfr}(b1)). 
The second event also occurs before the start of the MFR formation, 
during which a positive polarity patch enclosed by N2 and N3 disappears gradually (Figure~\ref{fig:bz_cancel}(e)-(h)). 
The NLFFFs reveal that this polarity connects to N3 initially (see Figure~\ref{fig:bf_mfr}(c)), 
thus the cancellation should occur between the positive polarity and N2, 
which results in some field lines from N3 connecting to P2 directly (see Figure~\ref{fig:bf_mfr}(d)), 
without significant separation occurring between P3 and N3. 
This event does not occur at the two identified BPs, 
but does occur at part of the identified collisional PIL. 
During the third event,  
a weak field region near N1 shrinks and disappears gradually (Figure~\ref{fig:bz_cancel}(i)-(l)). 
The process corresponds well with the continuous collisional shearing between N1 and the nearby mixed positive polarities, and the merging that creates the MFR there (see Figure~\ref{fig:fr_cmp}). 
Clear flux cancellation confirms the presence of magnetic reconnection in the very lower atmosphere through which the MFR is formed. 

\begin{figure*}
\begin{center}
\epsscale{1.1}
\plotone{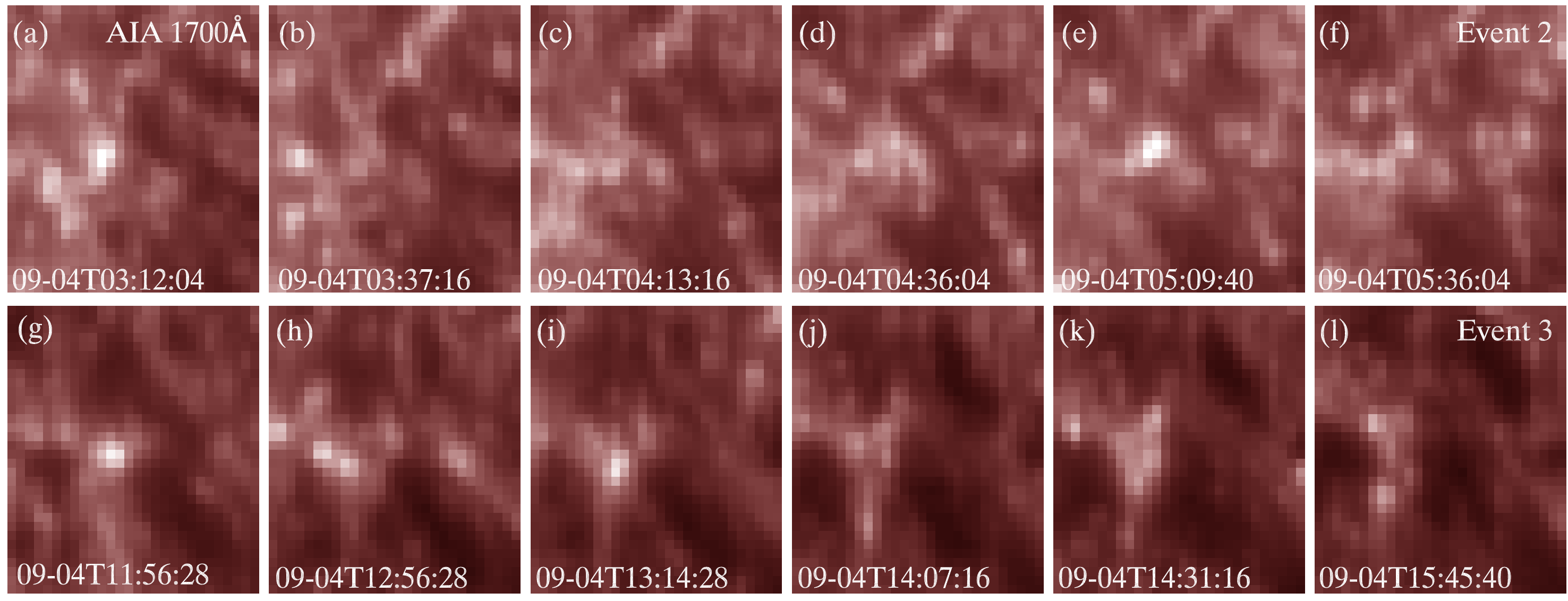}
\caption{Brightenings in the AIA 1700~\AA~passpand~occurring at the cancellation region   
during the Event2 and Event3 shown in Figure~\ref{fig:bz_cancel}. 
An online animation is also available.}\label{fig:bri_1700}
\end{center}
\end{figure*}

During the second and the third cancellation events, 
small-scale brightenings are observed in the AIA 1700~\AA~passband at the cancellation locations 
(Figure~\ref{fig:bri_1700}), again supporting the presence of magnetic reconnection. 
The brightenings in 1600~\AA~are also observed in the third event, but not as evident as the ones in 1700 \AA~(not shown here). 
These confirm that the reconnection mainly occurs in or slightly above the photosphere,  
since the 1700~\AA~passband emission usually originates from the temperature minimum region and the photosphere,  
while the 1600~\AA~emission usually originates from the upper photosphere and the transition region. 
For the first cancellation event, no evident UV brightenings are observed, 
which may be attributed to multiple reasons, such as a small amount of canceled flux, 
too weak reconnection to generate detectable brightenings, 
a reconnection site lower than the formation height of 1700~\AA~or 1600~\AA, 
or even combination of these factors.


\section{SUMMARY AND DISCUSSION}\label{sec:conc}

In this work, we investigate the formation of an 
MFR above the central PIL of NOAA AR 12673 during the AR's early emergence phase, mainly on 2017 September 4, 
through analyzing its photospheric magnetic field, coronal appearance imaged by the EUV and UV passbands, 
coronal field reconstructed by the NLFFFs model, 
and the photospheric motions calculated by the DAVE4VM method. 
It is found that when several bipoles emerge in the central region, 
their conjugated polarities separate from each other, 
resulting in collision between the non-conjugated opposite polarities at two locations. 
Nearly-potential loops connecting different bipoles appear at first, 
then seemingly get sheared and merge at the collision locations 
as evidenced by the appearance of a continuous sigmoidal filament in the 304~\AA~and a hot channel in the 131~\AA~passband, 
which also indicates the formation of an MFR.  
The NLFFFs further reveal that the MFR is built up gradually. 
It finally has relatively steday axial flux and average twist number of around $2.1\times10^{20}$~Mx and $-1.5$, respective. 
At its bottom, two elongated BPs (at the collision locations) and a slightly higher HFT are formed. 
Shearing motions are found developing between the non-conjugated polarities when the collision occurs. 
This process, 
named as ``collisional shearing", 
displays well correlation to the timings of the MFR formation. 
We also identify flux cancellation and UV brightenings at the BPs during the process. 

Combing the results, we conclude that the MFR is formed by the ``collisional shearing", specifically, 
through shearing and flux cancellation driven by the collision.   
The emerged magnetic field lines are sheared by the former and transferred into the MFR by the latter. 
Note that, 
we also identify collision between identical polarities, i.e., between P5 and the positive polarities near BP1, 
which does not drive direct reconnection, 
but does push the already collided opposite polarities to reconnect faster along the collisional PIL 
as evidenced by the faster increasing axial flux after the collision, speeding up the MFR formation. 
It provides another solid evidence to the critical role that collision plays in forming the MFR.

\cite{Sun_2018a} also identified the shearing motions, extending BPs and formation of a low-lying MFR above the central PIL prior to the two X-class flares on 2017 September 6.  
The MFR studied here have erupted during the CMEs on 2017 September 5~\citep{Vemareddy_2019} before its reformation reported in \cite{Sun_2018a}. 
These indicate that the MFR is reformed and erupted repeatedly,  
driven by the ``collisional shearing" resulting from the continuous flux emergence. 
It is consistent well with the findings in~\citet{Chintzoglou_2018}, 
which address the ``collisional shearing" 
as the main driver of major solar activities in emerging multipolar ARs.

The MFR here is proved to be 
formed through the interaction between different bipoles, 
thus the bodily emergence scenario, 
which expects the helical field lines emerge as a whole, is naturally excluded. 
The inconsistency between the bodily emergence scenario and the multipolar configurations was indeed originally suggested by~\citet{Chintzoglou_2018}.  
The sub-surface structure of the emerging magnetic field remains unknown. 
One may argue that the bipoles emerge with their own timings and flux contents, 
may suggest that they are independent. 
However, the four of the bipoles (except the biple N5 P5) are located side by side along the PIL, 
may be argued to evidence a possible sub-surface connection. 
It is also suggested by the simulation works 
that the different parts of one flux tube may rise at different times~\cite[e.g.,][]{Archontis_2008a, Archontis_2014, Leakej_2013, Leake_2014},
even with imbalanced flux~\citep{Fan_1999}. 
Thus, no conclusion can be made on the sub-surface configuration 
based on the present results.

As clear cancellation is observed, 
the reconnection that forms the MFR should 
mainly take place at, or slightly above, the photosphere as required by the cancellation model~\citep{VanBallegooijen_1989}. 
It is further evidenced by the UV brightenings, 
and supported by the successive appearance of a sigmoidal filament in the AIA 304 \AA~passband and a hot channel in the 131 \AA~passband. 
Note that transient EUV brightenings are also observed occasionally, 
but its association to the MFR formation seems hard to be determined. 
Even though a slightly higher HFT appears,
where the reconnection may also occur and play a role in forming and heating the MFR, 
it still does not reach the corona.

To summarize, 
we find that during the early emergence phase of NOAA AR 12673, the MFR above its central PIL is formed in the very lower atmosphere, through shearing and flux cancellation driven by the collision between emerging non-conjugated polarities, i.e., a process named as ``collisional shearing".  
The collision indeed results from the proper separations of the emerging conjugated polarities.

\clearpage

\clearpage
\bibliographystyle{aasjournal} 
\bibliography{MFR_EVO_12673}

\begin{thebibliography}{}
\expandafter\ifx\csname natexlab\endcsname\relax\def\natexlab#1{#1}\fi

\bibitem[{Amari {et~al.}(2010)Amari, Aly, Mikic, \& Linker}]{Amari_2010a}
Amari, T., Aly, J.-J., Mikic, Z., \& Linker, J. 2010, The Astrophysical
  Journal, 717, L26

\bibitem[{Amari {et~al.}(2004)Amari, Luciani, \& Aly}]{Amari_2004}
Amari, T., Luciani, J.~F., \& Aly, J.~J. 2004, The Astrophysical Journal, 615,
  L165

\bibitem[{Amari {et~al.}(2005)Amari, Luciani, \& Aly}]{Amari_2005}
---. 2005, The Astrophysical Journal, 629, L37

\bibitem[{Amari {et~al.}(2003{\natexlab{a}})Amari, Luciani, Aly, Mikic, \&
  Linker}]{Amari_2003a}
Amari, T., Luciani, J.~F., Aly, J.~J., Mikic, Z., \& Linker, J.
  2003{\natexlab{a}}, The Astrophysical Journal, 585, 1073

\bibitem[{Amari {et~al.}(2003{\natexlab{b}})Amari, Luciani, Aly, Mikic, \&
  Linker}]{Amari_2003b}
---. 2003{\natexlab{b}}, The Astrophysical Journal, 595, 1231

\bibitem[{Antiochos {et~al.}(1999)Antiochos, DeVore, \&
  Klimchuk}]{Antiochos_1999}
Antiochos, S.~K., DeVore, C.~R., \& Klimchuk, J.~a. 1999, The Astrophysical
  Journal, 510, 485

\bibitem[{Archontis \& Hood(2010)}]{Archontis_2010}
Archontis, V., \& Hood, A.~W. 2010, Astronomy and Astrophysics, 514, A56

\bibitem[{Archontis \& Hood(2012)}]{Archontis_2012}
---. 2012, Astronomy {\&} Astrophysics, 537, A62

\bibitem[{Archontis {et~al.}(2014)Archontis, Hood, \&
  Tsinganos}]{Archontis_2014}
Archontis, V., Hood, A.~W., \& Tsinganos, K. 2014, Astrophysical Journal
  Letters, 786, 2

\bibitem[{Archontis {et~al.}(2004)Archontis, Moreno-Insertis, Galsgaard, Hood,
  \& O'Shea}]{Archontis_2004}
Archontis, V., Moreno-Insertis, F., Galsgaard, K., Hood, A., \& O'Shea, E.
  2004, Astronomy {\&} Astrophysics, 426, 1047

\bibitem[{Archontis \& T{\"{o}}r{\"{o}}k(2008)}]{Archontis_2008a}
Archontis, V., \& T{\"{o}}r{\"{o}}k, T. 2008, Astronomy {\&} Astrophysics, 492,
  L35

\bibitem[{Aulanier {et~al.}(2010)Aulanier, T{\"{o}}r{\"{o}}k, D{\'{e}}moulin,
  \& DeLuca}]{Aulanier_torok_2010}
Aulanier, G., T{\"{o}}r{\"{o}}k, T., D{\'{e}}moulin, P., \& DeLuca, E.~E. 2010,
  The Astrophysical Journal, 708, 314

\bibitem[{Berger \& Field(1984)}]{Berger_Field_1984}
Berger, M.~a., \& Field, G.~B. 1984, Journal of Fluid Mechanics, 147, 133

\bibitem[{Bernasconi {et~al.}(2002)Bernasconi, Rust, Georgoulis, \&
  Labonte}]{Bernasconi_2002}
Bernasconi, P.~N., Rust, D.~M., Georgoulis, M.~K., \& Labonte, B.~J. 2002,
  Solar Physics, 209, 119

\bibitem[{Bobra {et~al.}(2014)Bobra, Sun, Hoeksema, Turmon, Liu, Hayashi,
  Barnes, \& Leka}]{Bobra_2014}
Bobra, M.~G., Sun, X., Hoeksema, J.~T., {et~al.} 2014, Solar Physics, 289, 3549

\bibitem[{Chen(2017)}]{Chenj_2017}
Chen, J. 2017, Physics of Plasmas, 24, 090501

\bibitem[{Cheng {et~al.}(2014)Cheng, Ding, Zhang, Sun, Guo, Wang, Kliem, \&
  Deng}]{Cheng_2014a}
Cheng, X., Ding, M.~D., Zhang, J., {et~al.} 2014, The Astrophysical Journal,
  789, 93

\bibitem[{Cheng {et~al.}(2017)Cheng, Guo, \& Ding}]{Chengx_2017}
Cheng, X., Guo, Y., \& Ding, M. 2017, Science China Earth Sciences, 60, 1383

\bibitem[{Cheung \& Isobe(2014)}]{Cheung_2014}
Cheung, M. C.~M., \& Isobe, H. 2014, Living Reviews in Solar Physics, 11,
  doi:10.12942/lrsp-2014-3

\bibitem[{Cheung {et~al.}(2008)Cheung, Sch{\"{u}}ssler, Tarbell, \&
  Title}]{Cheung_2008}
Cheung, M. C.~M., Sch{\"{u}}ssler, M., Tarbell, T.~D., \& Title, A.~M. 2008,
  The Astrophysical Journal, 687, 1373

\bibitem[{Chintzoglou {et~al.}(2015)Chintzoglou, Patsourakos, \&
  Vourlidas}]{Chintzoglou_2015}
Chintzoglou, G., Patsourakos, S., \& Vourlidas, A. 2015, Astrophys. J., 809, 34

\bibitem[{Chintzoglou {et~al.}(2019)Chintzoglou, Zhang, Cheung, \&
  Kazachenko}]{Chintzoglou_2018}
Chintzoglou, G., Zhang, J., Cheung, M. C.~M., \& Kazachenko, M. 2019, The
  Astrophysical Journal, 871, 67

\bibitem[{D{\'{e}}moulin(2006)}]{Demoulin_2006}
D{\'{e}}moulin, P. 2006, Advances in Space Research, 37, 1269

\bibitem[{D{\'{e}}moulin {et~al.}(1997)D{\'{e}}moulin, Bagala, Mandrini,
  Henoux, \& Rovira}]{Demoulin_1997}
D{\'{e}}moulin, P., Bagala, L.~G., Mandrini, C.~H., Henoux, J.~C., \& Rovira,
  M.~G. 1997, Astronomy and Astrophysics, 325, 305

\bibitem[{D{\'{e}}moulin {et~al.}(1996)D{\'{e}}moulin, H{\'{e}}noux, Priest, \&
  Mandrini}]{Demoulin_1996}
D{\'{e}}moulin, P., H{\'{e}}noux, J.~C., Priest, E.~R., \& Mandrini, C.~H.
  1996, Astronomy and Astrophysics, 308, 643

\bibitem[{DeVore(2000)}]{Devore_2000}
DeVore, C.~R. 2000, The Astrophysical Journal, 539, 944

\bibitem[{Emonet \& Moreno‐Insertis(1998)}]{Emonet_1998}
Emonet, T., \& Moreno‐Insertis, F. 1998, The Astrophysical Journal, 492, 804

\bibitem[{Fan(2001)}]{Fanyh_2001}
Fan, Y. 2001, The Astrophysical Journal, 554, L111

\bibitem[{Fan(2009)}]{Fanyh_2009}
---. 2009, The Astrophysical Journal, 697, 1529

\bibitem[{Fan(2012)}]{Fan_2012}
---. 2012, The Astrophysical Journal, 758, 60

\bibitem[{Fan \& Gibson(2004)}]{Fan_2004}
Fan, Y., \& Gibson, S.~E. 2004, The Astrophysical Journal, 609, 1123

\bibitem[{Fan {et~al.}(1999)Fan, Zweibel, Linton, \& Fisher}]{Fan_1999}
Fan, Y., Zweibel, E.~G., Linton, M.~G., \& Fisher, G.~H. 1999, The
  Astrophysical Journal, 521, 460

\bibitem[{Green \& Kliem(2009)}]{Green_2009}
Green, L.~M., \& Kliem, B. 2009, The Astrophysical Journal, 700, L83

\bibitem[{Green {et~al.}(2011)Green, Kliem, \& Wallace}]{Green_2011}
Green, L.~M., Kliem, B., \& Wallace, A.~J. 2011, Astronomy {\&} Astrophysics,
  526, A2

\bibitem[{Guo {et~al.}(2013)Guo, Ding, Cheng, Zhao, \& Pariat}]{Guoy_2013}
Guo, Y., Ding, M.~D., Cheng, X., Zhao, J., \& Pariat, E. 2013, The
  Astrophysical Journal, 779, 157

\bibitem[{Hoeksema {et~al.}(2014)Hoeksema, Liu, Hayashi, Sun, Schou, Couvidat,
  Norton, Bobra, Centeno, Leka, Barnes, Turmon, \& Others}]{Hoeksema_etal_2014}
Hoeksema, J.~T., Liu, Y., Hayashi, K., {et~al.} 2014, Solar Physics, 289, 3483

\bibitem[{Hou {et~al.}(2018)Hou, Zhang, Li, Yang, \& Li}]{Hou_2018}
Hou, Y.~J., Zhang, J., Li, T., Yang, S.~H., \& Li, X.~H. 2018, Astronomy {\&}
  Astrophysics, 1

\bibitem[{Inoue {et~al.}(2018)Inoue, Shiota, Bamba, \& Park}]{Inoue_2018a}
Inoue, S., Shiota, D., Bamba, Y., \& Park, S.-H. 2018, The Astrophysical
  Journal, 867, 83

\bibitem[{James {et~al.}(2017)James, Green, Palmerio, Valori, Reid, Baker,
  Brooks, van Driel-Gesztelyi, \& Kilpua}]{James_2017}
James, A.~W., Green, L.~M., Palmerio, E., {et~al.} 2017, Solar Physics, 292, 71

\bibitem[{Jiang {et~al.}(2014)Jiang, Wu, Feng, \& Hu}]{Jiang_2014a}
Jiang, C., Wu, S.~T., Feng, X., \& Hu, Q. 2014, The Astrophysical Journal, 780,
  55

\bibitem[{Joshi {et~al.}(2014)Joshi, Magara, \& Inoue}]{Joshi_2014a}
Joshi, N.~C., Magara, T., \& Inoue, S. 2014, Astrophysical Journal, 795,
  doi:10.1088/0004-637X/795/1/4

\bibitem[{Jouve \& Brun(2009)}]{Jouve_2009}
Jouve, L., \& Brun, A.~S. 2009, Astrophysical Journal, 701, 1300

\bibitem[{Kliem \& T{\"{o}}r{\"{o}}k(2006)}]{Kliem_2006}
Kliem, B., \& T{\"{o}}r{\"{o}}k, T. 2006, Physical Review Letters, 96, 255002

\bibitem[{Kuckein {et~al.}(2012{\natexlab{a}})Kuckein, {Mart{\'{i}}nez Pillet},
  \& Centeno}]{Kuckein_2012a}
Kuckein, C., {Mart{\'{i}}nez Pillet}, V., \& Centeno, R. 2012{\natexlab{a}},
  Astronomy {\&} Astrophysics, 539, A131

\bibitem[{Kuckein {et~al.}(2012{\natexlab{b}})Kuckein, {Mart{\'{i}}nez Pillet},
  \& Centeno}]{Kuckein_2012b}
---. 2012{\natexlab{b}}, Astronomy {\&} Astrophysics, 542, A112

\bibitem[{Leake {et~al.}(2014)Leake, Linton, \& Antiochos}]{Leake_2014}
Leake, J.~E., Linton, M.~G., \& Antiochos, S.~K. 2014, Astrophysical Journal,
  787, arXiv:1402.2645

\bibitem[{Leake {et~al.}(2013)Leake, Linton, \&
  T{\"{o}}r{\"{o}}k}]{Leakej_2013}
Leake, J.~E., Linton, M.~G., \& T{\"{o}}r{\"{o}}k, T. 2013, The Astrophysical
  Journal, 778, 99

\bibitem[{Lemen {et~al.}(2012)Lemen, Title, Akin, Boerner, Chou, Drake, Duncan,
  Edwards, Friedlaender, Heyman, Hurlburt, Katz, Kushner, Levay, Lindgren,
  Mathur, McFeaters, Mitchell, Rehse, Schrijver, Springer, Stern, Tarbell,
  Wuelser, Wolfson, Yanari, Bookbinder, Cheimets, Caldwell, Deluca, Gates,
  Golub, Park, Podgorski, Bush, Scherrer, Gummin, Smith, Auker, Jerram, Pool,
  Soufli, Windt, Beardsley, Clapp, Lang, \& Waltham}]{Lemen_etal_2012}
Lemen, J.~R., Title, A.~M., Akin, D.~J., {et~al.} 2012, Solar Physics, 275, 17

\bibitem[{Lites {et~al.}(1995)Lites, Low, {Martinez Pillet}, Seagraves,
  Skumanich, Frank, Shine, \& Tsuneta}]{Lites_1995}
Lites, B.~W., Low, B.~C., {Martinez Pillet}, V., {et~al.} 1995, The
  Astrophysical Journal, 446, 877

\bibitem[{Lites {et~al.}(2010)Lites, Kubo, Berger, Frank, Shine, Tarbell,
  Title, Okamoto, \& Otsuji}]{Lites_2010}
Lites, B.~W., Kubo, M., Berger, T., {et~al.} 2010, Astrophysical Journal, 718,
  474

\bibitem[{Liu {et~al.}(2018)Liu, Cheng, Wang, Zhou, Guo, \& Cui}]{Liu_2018d}
Liu, L., Cheng, X., Wang, Y., {et~al.} 2018, The Astrophysical Journal Letters,
  867, L5

\bibitem[{Liu {et~al.}(2017)Liu, Wang, Liu, Zhou, Temmer, Thalmann, Liu, Liu,
  Shen, Zhang, \& Veronig}]{Lliu_2017}
Liu, L., Wang, Y., Liu, R., {et~al.} 2017, The Astrophysical Journal, 844, 141

\bibitem[{Liu {et~al.}(2016)Liu, Kliem, Titov, Chen, Wang, Wang, Liu, Xu, \&
  Wiegelmann}]{rliu_2016}
Liu, R., Kliem, B., Titov, V.~S., {et~al.} 2016, The Astrophysical Journal,
  818, 148

\bibitem[{Longcope \& Welsch(2000)}]{Longcope_2000}
Longcope, D.~W., \& Welsch, B.~T. 2000, The Astrophysical Journal, 545, 1089

\bibitem[{Low(2001)}]{Blow_2001}
Low, B.~C. 2001, Journal of Geophysical Research, 106, 25,141

\bibitem[{Mackay {et~al.}(2010)Mackay, Karpen, Ballester, Schmieder, \&
  Aulanier}]{Mackay_2010}
Mackay, D.~H., Karpen, J.~T., Ballester, J.~L., Schmieder, B., \& Aulanier, G.
  2010, Space Science Reviews, 151, 333

\bibitem[{Mackay \& van Ballegooijen(2006)}]{Mackay_2006}
Mackay, D.~H., \& van Ballegooijen, a.~a. 2006, The Astrophysical Journal, 641,
  577

\bibitem[{{Manchester IV} {et~al.}(2004){Manchester IV}, Gombosi, DeZeeuw, \&
  Fan}]{ManchesterIV_2004}
{Manchester IV}, W., Gombosi, T., DeZeeuw, D., \& Fan, Y. 2004, The
  Astrophysical Journal, 610, 588

\bibitem[{Martens \& Zwaan(2002)}]{Martens_2002}
Martens, P.~C., \& Zwaan, C. 2002, The Astrophysical Journal, 558, 872

\bibitem[{Mitra {et~al.}(2018)Mitra, Joshi, Prasad, Veronig, \&
  Bhattacharyya}]{Mitra_2018a}
Mitra, P.~K., Joshi, B., Prasad, A., Veronig, A.~M., \& Bhattacharyya, R. 2018,
  eprint arXiv:1810.13146, 869, 69

\bibitem[{Moore {et~al.}(2001)Moore, Sterling, Hudson, \& Lemen}]{Moore_2001}
Moore, R.~L., Sterling, A.~C., Hudson, H.~S., \& Lemen, J.~R. 2001, The
  Astrophysical Journal, 552, 833

\bibitem[{Morgan \& Druckm{\"{u}}ller(2014)}]{Morgan_2014a}
Morgan, H., \& Druckm{\"{u}}ller, M. 2014, Solar Physics, 289, 2945

\bibitem[{Murray {et~al.}(2006)Murray, Moreno-Insertis, Archontis, Hood, \&
  Galsgaard}]{Murray_2006}
Murray, M.~J., Moreno-Insertis, F., Archontis, V., Hood, A.~W., \& Galsgaard,
  K. 2006, Astronomy {\&} Astrophysics, 460, 909

\bibitem[{Okamoto {et~al.}(2008)Okamoto, Tsuneta, Lites, Kubo, Yokoyama,
  Berger, Ichimoto, Katsukawa, Nagata, Shibata, Shimizu, Shine, Suematsu,
  Tarbell, \& Title}]{Okamoto_2008}
Okamoto, T.~J., Tsuneta, S., Lites, B.~W., {et~al.} 2008, The Astrophysical
  Journal, 673, L215

\bibitem[{Okamoto {et~al.}(2009)Okamoto, Tsuneta, Lites, Kubo, Yokoyama,
  Berger, Ichimoto, Katsukawa, Nagata, Shibata, Shimizu, Shine, Suematsu,
  Tarbell, \& Title}]{Okamoto_2009}
---. 2009, The Astrophysical Journal, 697, 913

\bibitem[{Pariat {et~al.}(2004)Pariat, Aulanier, Schmieder, Georgoulis, Rust,
  \& Bernasconi}]{Pariat_2004}
Pariat, E., Aulanier, G., Schmieder, B., {et~al.} 2004, The Astrophysical
  Journal, 614, 1099

\bibitem[{Pariat {et~al.}(2009)Pariat, Masson, \& Aulanier}]{Pariat_2009a}
Pariat, E., Masson, S., \& Aulanier, G. 2009, The Astrophysical Journal, 701,
  1911

\bibitem[{Patsourakos {et~al.}(2013)Patsourakos, Vourlidas, \&
  Stenborg}]{Patsourakos_2013}
Patsourakos, S., Vourlidas, A., \& Stenborg, G. 2013, The Astrophysical
  Journal, 764, 125

\bibitem[{Pesnell {et~al.}(2012)Pesnell, Thompson, \&
  Chamberlin}]{Pesnell_2012}
Pesnell, W.~D., Thompson, B.~J., \& Chamberlin, P.~C. 2012, Solar Physics, 275,
  3

\bibitem[{Romano {et~al.}(2018)Romano, Elmhamdi, Falco, Costa, Kordi,
  Al-Trabulsy, \& Al-Shammari}]{Romano_2018}
Romano, P., Elmhamdi, A., Falco, M., {et~al.} 2018, The Astrophysical Journal,
  852, L10

\bibitem[{Savcheva {et~al.}(2012)Savcheva, Green, {Van Ballegooijen}, \&
  Deluca}]{Savcheva_2012}
Savcheva, A.~S., Green, L.~M., {Van Ballegooijen}, A.~A., \& Deluca, E.~E.
  2012, Astrophysical Journal, 759, doi:10.1088/0004-637X/759/2/105

\bibitem[{Schmieder {et~al.}(2004)Schmieder, Mein, Deng, Dumitrache, Malherbe,
  Staiger, \& Deluca}]{Schmieder_2004b}
Schmieder, B., Mein, N., Deng, Y., {et~al.} 2004, Solar Physics, 223, 119

\bibitem[{Schuck(2008)}]{Schuck_2008}
Schuck, P.~W. 2008, The Astrophysical Journal, 683, 1134

\bibitem[{Shen {et~al.}(2018)Shen, Xu, Wang, Chi, \& Luo}]{Shen_2018}
Shen, C., Xu, M., Wang, Y., Chi, Y., \& Luo, B. 2018, The Astrophysical
  Journal, 861, 28

\bibitem[{Shibata {et~al.}(1995)Shibata, Masuda, Shimojo, Hara, Yokoyama,
  Tsuneta, Kosugi, \& Ogawara}]{Shibata_1995}
Shibata, K., Masuda, S., Shimojo, M., {et~al.} 1995, The Astrophysical Journal,
  451, doi:10.1086/309688

\bibitem[{Sun {et~al.}(2018)Sun, Kazachenko, Titov, \& Cheung}]{Sun_2018a}
Sun, X., Kazachenko, M., Titov, V., \& Cheung, M. 2018, in 42nd COSPAR
  Scientific Assembly

\bibitem[{Sun \& Norton(2017)}]{Sunxd_rna_2017}
Sun, X., \& Norton, A.~A. 2017, Research Notes of the AAS, 1, 24

\bibitem[{Titov \& D{\'{e}}moulin(1999)}]{Titov_1999}
Titov, V.~S., \& D{\'{e}}moulin, P. 1999, ASTRONOMY AND ASTROPHYSICS, 351, 707

\bibitem[{Titov {et~al.}(2002)Titov, Hornig, \& D{\'{e}}moulin}]{Titov_2002}
Titov, V.~S., Hornig, G., \& D{\'{e}}moulin, P. 2002, Journal of Geophysical
  Research: Space Physics, 107, SSH 3

\bibitem[{Titov {et~al.}(1993)Titov, Priest, \& D{\'{e}}moulin}]{Titov_1993}
Titov, V.~S., Priest, E.~R., \& D{\'{e}}moulin, P. 1993, Astronomy and
  Astrophysics, 276, 564

\bibitem[{T{\"{o}}r{\"{o}}k {et~al.}(2004)T{\"{o}}r{\"{o}}k, Kliem, \&
  Titov}]{Torok_2004}
T{\"{o}}r{\"{o}}k, T., Kliem, B., \& Titov, V.~S. 2004, Astronomy and
  Astrophysics, 413, 27

\bibitem[{Valori {et~al.}(2012)Valori, Green, D{\'{e}}moulin, {Vargas
  Dom{\'{i}}nguez}, van Driel-Gesztelyi, Wallace, Baker, \&
  Fuhrmann}]{Valori_2012}
Valori, G., Green, L.~M., D{\'{e}}moulin, P., {et~al.} 2012, Solar Physics,
  278, 73

\bibitem[{van Ballegooijen \& Martens(1989)}]{VanBallegooijen_1989}
van Ballegooijen, A.~A., \& Martens, P. C.~H. 1989, The Astrophysical Journal,
  343, 971

\bibitem[{{Vargas Dom{\'{i}}nguez} {et~al.}(2012){Vargas Dom{\'{i}}nguez},
  MacTaggart, Green, van Driel-Gesztelyi, \& Hood}]{Vargas_2012}
{Vargas Dom{\'{i}}nguez}, S., MacTaggart, D., Green, L., van Driel-Gesztelyi,
  L., \& Hood, A.~W. 2012, Solar Physics, 278, 33

\bibitem[{Vemareddy(2019)}]{Vemareddy_2019}
Vemareddy, P. 2019, The Astrophysical Journal, 872, 182

\bibitem[{Verma(2018)}]{Verma_2018}
Verma, M. 2018, Astronomy {\&} Astrophysics, 612, A101

\bibitem[{Veronig {et~al.}(2018)Veronig, Podladchikova, Dissauer, Temmer, {B.
  Seaton}, Long, Guo, Vr{\v{s}}nak, Harra, \& Kliem}]{Veronig_2018}
Veronig, A.~M., Podladchikova, T., Dissauer, K., {et~al.} 2018, The
  Astrophysical Journal, 868, 107

\bibitem[{Wang {et~al.}(2018)Wang, Liu, Hoeksema, Zimovets, \&
  Liu}]{Wang_2018e}
Wang, R., Liu, Y.~D., Hoeksema, J.~T., Zimovets, I.~V., \& Liu, Y. 2018, The
  Astrophysical Journal, 869, 90

\bibitem[{Welsch {et~al.}(2005)Welsch, DeVore, \& Antiochos}]{Welsch_2005}
Welsch, B.~T., DeVore, C.~R., \& Antiochos, S.~K. 2005, The Astrophysical
  Journal, 634, 1395

\bibitem[{Wheatland {et~al.}(2000)Wheatland, Sturrock, \&
  Roumeliotis}]{Wheatland_2000}
Wheatland, M.~S., Sturrock, P.~a., \& Roumeliotis, G. 2000, The Astrophysical
  Journal, 540, 1150

\bibitem[{Wiegelmann(2004)}]{Wiegelmann_2004}
Wiegelmann, T. 2004, Solar Physics, 219, 87

\bibitem[{Wiegelmann \& Inhester(2010)}]{Wiegelmann_2010b}
Wiegelmann, T., \& Inhester, B. 2010, Astronomy and Astrophysics, 516, A107

\bibitem[{Wiegelmann {et~al.}(2006)Wiegelmann, Inhester, \&
  Sakurai}]{Wiegelmann_2006}
Wiegelmann, T., Inhester, B., \& Sakurai, T. 2006, Solar Physics, 233, 215

\bibitem[{Wiegelmann {et~al.}(2012)Wiegelmann, Thalmann, Inhester, Tadesse,
  Sun, \& Hoeksema}]{Wiegelmann_2012}
Wiegelmann, T., Thalmann, J.~K., Inhester, B., {et~al.} 2012, Solar Physics,
  281, 37

\bibitem[{Yan {et~al.}(2016)Yan, Priest, Guo, Xue, Wang, \& Yang}]{Yanxl_2016}
Yan, X.~L., Priest, E.~R., Guo, Q.~L., {et~al.} 2016, The Astrophysical
  Journal, 832, 1

\bibitem[{Yan {et~al.}(2018)Yan, Wang, Pan, Kong, Xue, Yang, Li, \&
  Feng}]{Yanxl_2018a}
Yan, X.~L., Wang, J.~C., Pan, G.~M., {et~al.} 2018, The Astrophysical Journal,
  856, 79

\bibitem[{Yan {et~al.}(2015)Yan, Xue, Pan, Wang, Xiang, Kong, \&
  Yang}]{Yanxl_2015}
Yan, X.~L., Xue, Z.~K., Pan, G.~M., {et~al.} 2015, The Astrophysical Journal
  Supplement Series, 219, 17

\bibitem[{Yang {et~al.}(2017)Yang, Zhang, Zhu, \& Song}]{Yangsh_2017}
Yang, S., Zhang, J., Zhu, X., \& Song, Q. 2017, The Astrophysical Journal, 849,
  L21

\bibitem[{Zou {et~al.}(2019)Zou, Jiang, Feng, Zuo, Wang, \& Wei}]{Zou_2019}
Zou, P., Jiang, C., Feng, X., {et~al.} 2019, The Astrophysical Journal, 870, 97

\bibitem[{Zwaan(1985)}]{Zwaan_1985}
Zwaan, C. 1985, Solar Physics, 100, 397

\end{thebibliography}

\acknowledgments{We thank Georgios Chintzoglou, Xudong Sun and Jun Chen for valuable discussions, 
and our anonymous referee for his/her constructive comments that significantly improved the manuscript. 
We acknowledge the use of data from {\it SDO} 
and computation support provided by National Supercomputer Center in Guangzhou through HPC resources on Tianhe-2. 
L.L. is supported by NSFC (11803096) and the Open Project of CAS Key Laboratory of Geospace Environment. 
X.C. is funded by NSFC (11722325, 11733003, 11790303, 11790300) and Jiangsu NSF (BK20170011). X.C. also thanks the Alexander von Humboldt foundation for supporting his stay in Germany. 
Y.W. is supported by NSFC (41574165, 41774178). 

\end{document}